\documentclass[aps,prl,nobibnotes,twocolumn,superscriptaddress,bibliography,floatfix]{revtex4-2}
\usepackage{amsfonts}
\usepackage{mathrsfs}
\usepackage{amsmath}
\usepackage{color}
\usepackage{graphicx}
\usepackage{bm}
\usepackage{amssymb}
\usepackage{xspace}
\usepackage{epstopdf}
\usepackage{dcolumn}
\usepackage{longtable}
\usepackage{multirow}
\usepackage{float}
\usepackage{comment}
\usepackage[normalem]{ulem}
\makeatletter

\usepackage[colorlinks=true, letterpaper=true, pdfstartview=FitV, linkcolor=blue, citecolor=blue, urlcolor=blue]{hyperref}

\makeatother

\begin{document}

\title{Fractional Spin Ferroelectric and Sliding Spin Current in  Magnetic Sliding Ferroelectrics}

\author{Yilin Han}
\affiliation{Key Lab of advanced optoelectronic quantum architecture and measurement (MOE), Beijing Key Laboratory of Quantum Matter State Control and Ultra-Precision Measurement Technology, and School of Physics, Beijing Institute of Technology, Beijing 100081, China}

\author{Lei Li}
\affiliation{Research Center for Quantum Physics and Technologies, Inner Mongolia University, Inner Mongolia Key Laboratory of Microscale Physics and Atomic Manufacturing, Inner Mongolia University,\\ and School of Physical Science and Technology, Inner Mongolia University, Hohhot 010021, China}

\author{Chaoxi Cui}
\affiliation{Key Lab of advanced optoelectronic quantum architecture and measurement (MOE), Beijing Key Laboratory of Quantum Matter State Control and Ultra-Precision Measurement Technology, and School of Physics, Beijing Institute of Technology, Beijing 100081, China}

\author{Run-Wu Zhang}
\affiliation{Key Lab of advanced optoelectronic quantum architecture and measurement (MOE), Beijing Key Laboratory of Quantum Matter State Control and Ultra-Precision Measurement Technology, and School of Physics, Beijing Institute of Technology, Beijing 100081, China}

\author{Zhi-Ming Yu}
\email{zhiming\_yu@bit.edu.cn}
\affiliation{Key Lab of advanced optoelectronic quantum architecture and measurement (MOE), Beijing Key Laboratory of Quantum Matter State Control and Ultra-Precision Measurement Technology, and School of Physics, Beijing Institute of Technology, Beijing 100081, China}
\author{Yugui Yao}
\affiliation{Key Lab of advanced optoelectronic quantum architecture and measurement (MOE), Beijing Key Laboratory of Quantum Matter State Control and Ultra-Precision Measurement Technology, and School of Physics, Beijing Institute of Technology, Beijing 100081, China}

\begin{abstract}
We investigate the fractional spin ferroelectric (FSFE) in  magnetic sliding ferroelectrics (SFEs), where ferroelectric switching is characterized not only by the reversal of the out-of-plane electric polarization but also by a variation of fractional in-plane spin  electronic polarization.
We show that interlayer sliding in FSFEs can naturally lead to a symmetry-protected pure spin current, termed the sliding spin current here.
The underlying mechanism is that, during switching, the contributions of valence electrons and ions to the in-plane charge transfer cancel each other, whereas the in-plane spin transfer, which stems solely from valence electrons, persists, leading to a pure spin current.
We demonstrate our ideas in  various  material candidates, including $H$-stacked bilayer CrI$_3$, whose few-layer form has been experimentally confirmed to be a magnetic SFE, and $R$-stacked bilayers $2H$-V$X_2$ ($X=$ S, Se, Te), which have been experimentally synthesised.
For a typical switching time of about $1$ ns, the estimated spin-current densities for bilayer CrI$_3$ and  V$X_2$ reach $10^9 (\hbar/2e)\mathrm{A/m^2}$  and  $10^8 (\hbar/2e)\mathrm{A/m^2}$, respectively.
This means that by applying a periodic out-of-plane electric field, a significant alternating spin current can be generated in magnetic SFEs.
Thus, our findings propose a compelling new mechanism for the all-electrical generation of pure spin current, and predict concrete realistic materials for experimental verification.
\end{abstract}

\maketitle

{\emph{\textcolor{blue}{Introduction.--}}}
Spintronics aims to manipulate information through spin degrees of freedom rather than charge alone, making the generation and control of spin current a central task in the field \cite{zutic2004spintronicsb,fert2008nobel,bader2010spintronicsb,hirohata2020reviewa}.
Over the past decades, several mechanisms have been proposed and have greatly advanced the research \cite{hirsch1999spin,tserkovnyak2005nonlocal,lee2021efficient,bose2022tilted,kikkawa2023spin,ding2024orbital,chen2025generation,zhang2025electrical,wang2025intrinsic}. For example, the spin Hall effect (SHE) converts a longitudinal charge current into a transverse spin current via spin-orbit coupling (SOC) \cite{hirsch1999spin,sinova2015spinb,kato2004observation,valenzuela2006direct,jungwirth2012spin,liu2012spintorque,zhu2019variationa,abdelwahab2024twodimensional,wang2025intrinsica}, and the ferromagnetic or antiferromagnetic spin pumping injects spin angular momentum from a magnetic system into a normal metal via magnetization dynamics \cite{tserkovnyak2005nonlocal,baltz2018antiferromagnetica,tserkovnyak2002enhanced,mosendz2010quantifying,cheng2014spin,hamara2024ultrahigh,tang2024thermal}.
However, a fully electrical generation of pure spin current without charge transport is highly desirable compared with SHE and magnetically (optically) driven mechanisms. Moreover, the SHE usually requires materials composed of heavy elements with strong SOC, which imposes stringent constraints on material candidates.

Meanwhile, unconventional ferroelectrics including sliding ferroelectrics (SFEs) \cite{li2017binaryd,wu2021slidinga,viznerstern2021interfacial,qi2021review,viznerstern2025sliding,zhang2025emerging} and fractional quantum ferroelectrics (FQFEs) \cite{ji2024fractionala,luo2026unified} have been attracting broad interest.
Two-dimensional SFEs have two distinct ground states with opposite out-of-plane electric polarizations ($\pm P_z$) \cite{li2017binaryd}.
Thus, a vertical electric field can drive the SFE from a state with $P_z$ to a state with $-P_z$ via interlayer sliding, or vice versa.
In addition, FQFEs are characterized by switchable, quantized electric polarization differences between two ground states \cite{ji2024fractionala,luo2026unified}.
Research on SFEs and FQFEs has progressed rapidly. Various material candidates have been predicted \cite{li2017binaryd,fei2018ferroelectricb,yasuda2021stackingengineeredc,viznerstern2021interfacial,meng2022slidinga,wan2022roomtemperature,wang2022interfacial,
sui2023sliding,yang2024ferroelectric,bian2024developinga,yasuda2024ultrafasta,bai2025subnanoseconda,ji2023generalc,yang2023atypicala,chen2024stronga,guo2025slidingb,
ji2024fractionala,jiang20252d,pang2025generalized,yu2025symmetry,luo2026unified,dong2026fractionala}, and many of them have been experimentally confirmed \cite{fei2018ferroelectricb,viznerstern2021interfacial,yasuda2021stackingengineeredc,wang2022interfacial,wan2022roomtemperature,
meng2022slidinga,sui2023sliding,yang2024ferroelectric,bian2024developinga,yasuda2024ultrafasta,bai2025subnanoseconda,ji2024fractionala,jiang20252d}.
In particular, SFEs have been experimentally demonstrated to be fatigue resistant \cite{bian2024developinga,yasuda2024ultrafasta}, and interlayer sliding can be completed within 1 ns \cite{yasuda2024ultrafasta,bai2025subnanoseconda}.

In this work, we bring together three important but seemingly unrelated concepts--spin current, quantized polarization, and SFE--to realize a new all-electrical mechanism for generating pure spin current.
For many  magnetic SFEs, we find that during interlayer sliding, although the in-plane electric polarization of the system shows no net change, the  variation of the in-plane spin electric polarization, which is defined as the difference in the electric polarization between  opposite spins here, is governed by a fractional polarization quantum.
Consequently, an out-of-plane electric field can induce an interlayer sliding and then generate a significant pure spin current in magnetic SFEs.
We designate this class of magnetic SFEs as fractional spin ferroelectrics (FSFEs) and the resulting  spin current as sliding spin current.

\begin{figure}[t]
\includegraphics[width=0.48\textwidth]{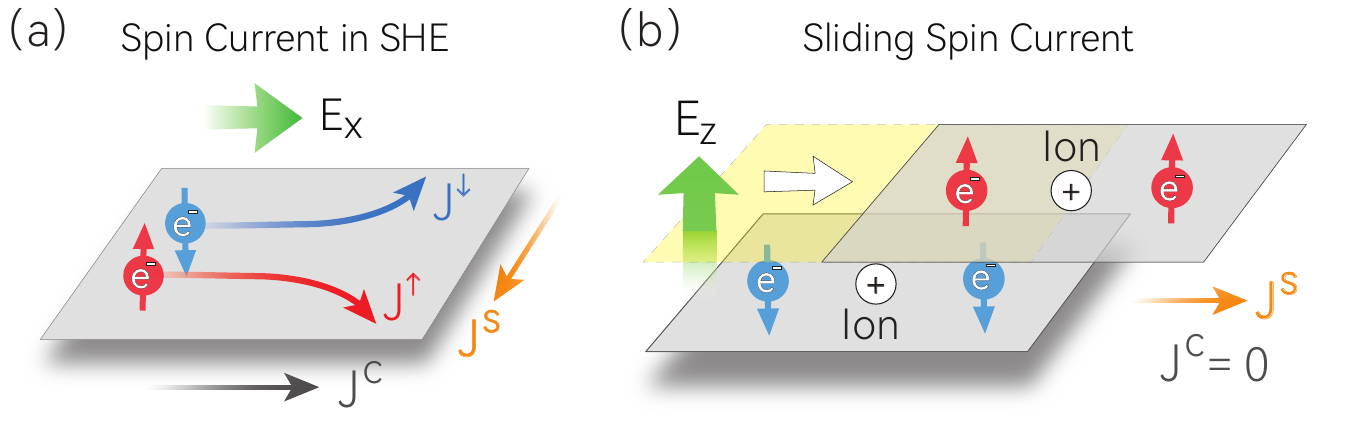}
\caption{\label{fig:mechanism} Schematics of pure spin current in (a)   SHE and  (b) sliding spin current.
In (b), since the ions--comprising atomic nuclei and core electrons--possess negligible magnetic moments, the simultaneous motion of valence electrons and ions in the top layer can lead to vanishing charge current but a significant spin current. }
\end{figure}

Via first-principles calculations, we confirm our idea in many realistic materials, including $H$-stacked bilayer CrI$_3$ and $R$-stacked bilayer $2H$-V$X_2$ ($X=$ S, Se, Te).
Moreover,  our calculations reveal that the underlying physics of the sliding spin current is fundamentally different from that of previous mechanisms.
For example, in the SHE, the pure spin current results from electrons with opposite spins moving in opposite directions [see Fig.~\ref{fig:mechanism}(a)].
In contrast, for sliding spin current, the pure spin current is guaranteed by the compensation of valence electrons and ions, which  propagate in the same direction during interlayer sliding, as illustrated in Fig.~\ref{fig:mechanism}(b).

Remarkably, while a static out-of-plane electric field can induce only a transient sliding spin current during ferroelectric switching,  a periodic electric field can drive repeated switching and thereby generate an alternating spin current. More importantly, the fast, nonvolatile, and fatigue-resistant electric-field switching of sliding ferroelectrics \cite{bian2024developinga,wang2022interfacial,yasuda2024ultrafasta,bai2025subnanoseconda} makes such alternating spin currents highly desirable for  spintronic functionalities.
For a typical switching time of about $1$ ns  \cite{yasuda2024ultrafasta,bai2025subnanoseconda}, our estimated spin-current density can reach $10^9 (\hbar/2e)\mathrm{A/m^2}$ for bilayer CrI$_3$ and $10^8 (\hbar/2e)\mathrm{A/m^2}$ for bilayer V$X_2$, which are comparable with the spin-current density generated by the SHE in heavy metals when the longitudinal charge current density is about $10^{9} \mathrm{A/m^2}$ \cite{liu2012spintorque,zhu2019variationa}.
Clearly, the sliding spin current proposed here  neither requires nor entails such high charge-current densities.
Moreover, the sliding spin current can appear in systems with negligible SOC, such as bilayer V$X_2$. 

{\emph{\textcolor{blue}{Symmetry analysis of sliding spin current.--}}}
For simplicity, we assume the magnetic SFEs have spin U$(1)$ symmetry. The electric polarization of the system can then be divided into three parts:
\begin{align}\label{EP}
{\bm P}={\bm P}^{\rm ion}+{\bm P}^{\uparrow}+{\bm P}^{\downarrow},
\end{align}
where ${\bm P}^{\rm ion}=\frac{e}{S}\sum_{n} Z_n {\bm \tau}_n \pmod{{\bm Q}}$ denotes the ionic contribution to the polarization,  $-e$ is  the electron charge,  $e Z_n$  (${\bm \tau}_n$) is the charge  (position) of the $n$th ion, and ${\bm Q}=e {\bm R}/S$ is the polarization quantum, with $S$ the unit-cell area and ${\bm R}$ a lattice vector \cite{vanderbilt2018berry}.
${\bm P}^{\uparrow(\downarrow)}=\frac{-e}{S}\sum_{n}^{occ}\int_{BZ} \frac{d^D k}{(2\pi)^D} {\bm A}_n^{\uparrow(\downarrow)}({\bm k}) \pmod{{\bm Q}}$ denotes the polarization contributed by valence electrons with up-spin (down-spin), where the summation runs over all occupied valence bands, $D$ is the dimensionality of the system, and ${\bm A}_n^{\uparrow(\downarrow)}$ is the Berry connection of the $n$th band in the up-spin (down-spin) channel.
Note that in the presence of additional crystalline symmetries, the polarization of the system is quantized, and can take only values of fractional polarization quantum \cite{ji2024fractionala,pang2025generalized,luo2026unified}.

Because the ions, which consist of atomic nuclei and core electrons, carry negligible net magnetic moment, spin transport in the system arises solely from the valence electrons. To describe the possible spin transport during ferroelectric switching, we define the spin electronic polarization as
\begin{align}\label{SP}
{\bm P}^s={\bm P}^{\uparrow}-{\bm P}^{\downarrow}.
\end{align}
The mean transient charge and spin currents during switching can then be expressed, respectively, as
\begin{align}\label{cur}
{\bm J}^{c}=\frac{\delta{\bm P}}{\Delta t}=\frac{{\cal O}{\bm P}-{\bm P}}{\Delta t}, \ \ \  {\bm J}^{s}=\frac{\delta{\bm P}^s}{\Delta t}=\frac{{\cal O}{\bm P}^s-{\bm P}^s}{\Delta t},
\end{align}
where $\Delta t$ is the switching time and ${\cal O}$ is the operator that connects the initial and final states of the target ferroelectric.
In general, only the point-group part of ${\cal O}$ is relevant to the symmetry analysis.

\begin{figure}[t]
\includegraphics[width=0.5\textwidth]{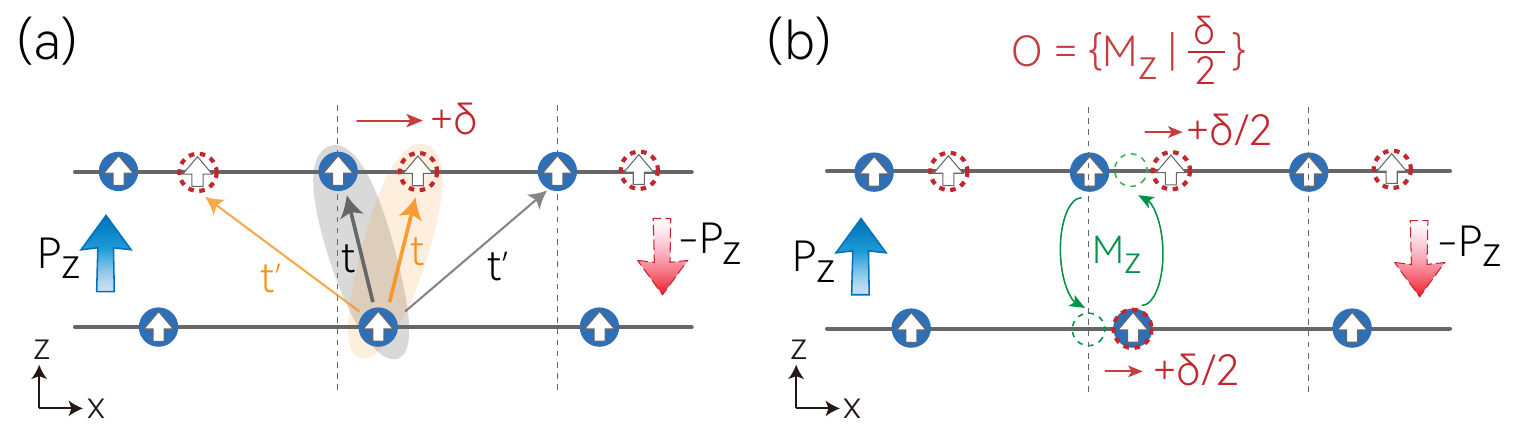}
\caption{\label{fig:model} A simple case illustrating the unique symmetry analysis of the sliding spin current.
(a) A ferromagnetic SFE, where the initial and final ferroelectric states are connected by sliding the top layer by a distance $\delta$.
(b) The  two ferroelectric states are  also  connected by a horizontal mirror followed by a $\delta/2$ translation  $\{{\cal M}_z|\delta/2\}$.
The blue and red arrows in (a-b) denote the electric polarization of the initial and final ferroelectric states, respectively.
}\end{figure}

Interestingly, the symmetry analysis for sliding spin current exhibits  unique features.
Consider a magnetic SFE without additional crystalline symmetry, as illustrated in Fig.~\ref{fig:model}(a), where the initial and final states are related by $\{E|\delta\}_{\rm top}$, which denotes sliding the top layer by a distance $\delta$.
However, $\{E|\delta\}_{\rm top}$  cannot be directly used for symmetry analysis.
Thus, we should identify another operator ${\cal O}$ that is suitable for symmetry analysis.
As illustrated in  Fig.~\ref{fig:model}(b), we  find that  the initial and final states of this simple case  are also connected by  ${\cal O}=\{{\cal M}_z|\frac{\delta}{2}\}$, a horizontal mirror followed by a $\delta/2$ translation of the entire system.
From $\{{\cal M}_z|\frac{\delta}{2}\}$, three important properties  can  be directly inferred.
(i) The  initial and final states of the magnetic SFE have opposite $P_z$, as guaranteed by ${\cal M}_z$.
(ii) Remarkably, the translation in ${\cal O}$ also affects $\delta{\bm P}$ and $\delta{\bm P}^s$.
$\{E|\frac{\delta}{2}\}$  indicates a  movement of the entire system, leading to a pure spin transport because the system is charge neutral but has a finite magnetic moment. (iii) More importantly, without any detailed calculation, one immediately finds that the variation of the spin electronic polarization is $\delta P_x^s=n^s e \frac{\delta}{2} \pmod{{\bm Q}}$ where $n^s=n^{\uparrow}-n^{\downarrow}$ is the difference between the numbers of up-spin and down-spin electrons. This result is not limited to the simple lattice in Fig.~\ref{fig:model} but applies to all magnetic SFEs with ${\cal O}=\{{\cal M}_z|\frac{\delta}{2}\}$.

Point (iii) can be confirmed numerically. For the lattice model defined in Fig.~\ref{fig:model},  the Hamiltonian of the initial (final) state  can be written as
\begin{align}
{\cal H}_{i(f)}=\left[\begin{array}{cc}
0 & t e^{\mp i k_x \delta/2}+t' e^{\pm i k_x \delta'}\\
t e^{\pm i k_x \delta/2}+t' e^{\mp i k_x \delta'} & 0
\end{array}\right],
\end{align}
where $t$ ($t'$) is the hopping between the top and bottom atoms, with  $|t|>|t'|$ [see Fig.~\ref{fig:model}(a)], and $\delta'=1-\delta/2$. Although the band structures of ${\cal H}_{i}$ and ${\cal H}_{f}$ are identical, the electric polarization of ${\cal H}_{i(f)}$ is $\mp e \delta/4$. Therefore, the variation $P_x^s$ is $e \delta/2$ for $n^s=1$ (corresponding to one  valence band), consistent with the symmetry analysis.
Adding additional symmetry-preserving hoppings, orbitals and atoms to ${\cal H}_{i(f)}$ do not alter the result obtained from the symmetry analysis.

{\emph{\textcolor{blue}{Conditions for FSFE and significant sliding spin current.--}}}
Generally, the variation of the spin electronic polarization $\delta {\bm P}^s$ in magnetic SFEs can be an arbitrary value, as illustrated in Fig.~\ref{fig:model}.
However, in the presence of additional symmetries, $\delta {\bm P}^s$  may be quantized to a fractional polarization quantum~\cite{chen2025generation}.

Surprisingly, we find that most currently known magnetic SFEs turn out to be FSFEs (see Table \ref{tab:table2}).
The quantized  $\delta {\bm P}^s$ results from both the point-group operator and the translation in ${\cal O}$, whose general form is  ${\cal O}=\{O|{\bm \tau}\}$.
To achieve a quantized $\delta {\bm P}^s$, two conditions should be satisfied.  (i)  $O{\bm P}^s-{\bm P}^s=\sum_{i}n_i {\bm F}_i$, where ${\bm F}_i$ is the $i$th fractional polarization quantum allowed by the symmetry of the system and $n_i$ is an integer.
For example, for a magnetic SFE with spatial inversion symmetry (${\cal I}$), there are two choices of fractional polarization quantum, i.e., ${\bm F}_1=0$ and ${\bm F}_2={\bm Q}/2$.
Without any crystalline symmetry, we have ${\bm F}_1=0$ and ${\bm F}_2={\bm Q}$.
(ii) ${\bm \tau}$ should be a fractional translation in units of the lattice vectors, which requires the magnetic atoms in the top and bottom layers of the magnetic SFE to be separated by a fractional translation.

A unique  and important feature of magnetic SFEs is that under the two aforementioned conditions, although the variation of the electronic polarization $\delta {\bm P}$ should also be quantized to fractional polarization quantum, i.e., $\delta {\bm P}=\sum_{i}n_i {\bm F}_i$,  it generally takes only the value zero: $\delta {\bm P}=0$.
This is because, owing to the vdW structure of SFEs, the net charge carried by each layer is generally very small (see Table \ref{tab:table2}).
Hence, interlayer sliding generally cannot generate a $\delta {\bm P}$ on the scale of ${\bm Q}$ or ${\bm F}_i$, and instead leads only to  $\delta {\bm P}=0$.
In sharp contrast, each layer of magnetic SFEs can exhibit a sizable magnetic moment, making  ${\bm P}^s$ finite.
Consequently, FSFEs can exist among magnetic SFEs, and interlayer sliding in an FSFE will generate a pure and significant spin current.

\begin{figure}[t]
\includegraphics[width=0.5\textwidth]{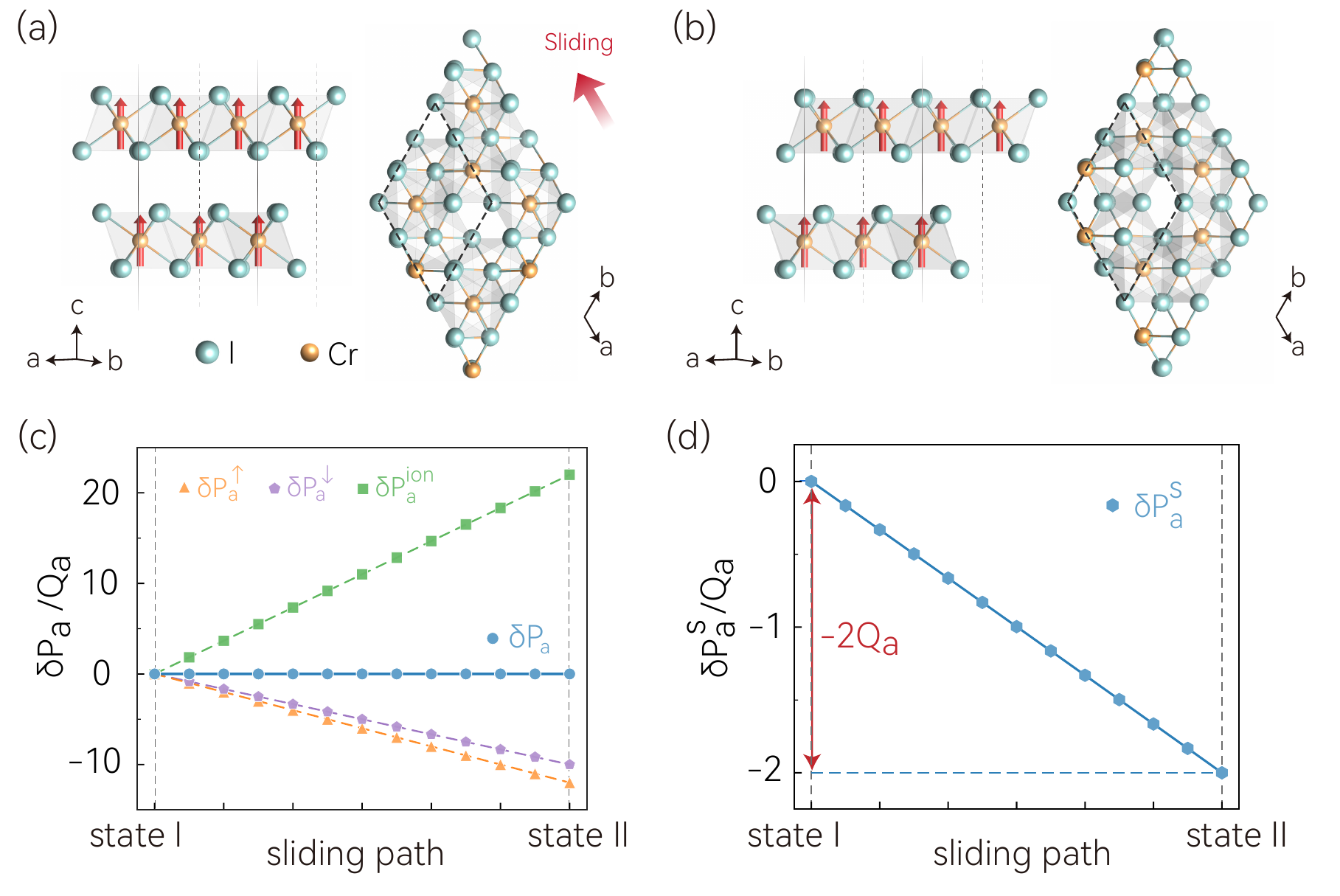}
\caption{\label{fig:CrI3} (a-b) Side and top views of $H$-stacked bilayer CrI$_3$ in ferroelectric (a) state I and (b) state II.
The bilayer CrI$_3$ can switch from state I to state II by shifting the top layer by $-\mathbf a/3$, as marked in (a).
(c-d) Variations of (c) polarizations and (d) spin electronic polarization along the (10) direction.  $Q_a=269.3$ pC/m denotes the (10) component of the polarization quantum.}
\end{figure}

{\emph{\textcolor{blue}{$H$-stacked bilayer CrI$_3$}.--}}
Recently, few-layer $H$-stacked CrI$_3$ has been experimentally demonstrated to be a magnetic SFE \cite{fox2025sliding}.
As a typical example, we study the sliding spin current in $H$-stacked bilayer CrI$_3$, whose ground state is interlayer ferromagnetic (FM).
Bilayer CrI$_3$ can be stabilized in two energetically degenerate configurations, denoted as state I and state II \cite{ji2023generalc}, which exhibit opposite out-of-plane polarizations of $P_z=\pm 0.1$ pC/m, as illustrated in Figs.~\ref{fig:CrI3}(a-b). Its magnetic space group is No. 8.34 and preserves either $M_{210}{\cal T}$ or $M_{1\bar10}{\cal T}$ symmetry. There are four Cr atoms and twelve I atoms in its magnetic unit cell. The magnetic moments are mainly localized on the four Cr atoms, each of which carries a moment of $3.35\mu_B$ along the $z$ direction.
In state I, the Cr atoms in the top and bottom layers reside at $(1/3,1/3)$ and $(2/3,0)$, and at $(2/3,1/3)$ and $(1/3,2/3)$, respectively, in fractional coordinates.
State I can be switched to state II by sliding the top atomic layer by $-\mathbf a/3$, as shown in  Figs.~\ref{fig:CrI3}(a-b).
The energy barrier of $H$-stacked bilayer CrI$_3$ along this sliding path is low \cite{ji2023generalc}.

With the bottom layer fixed,  the two ferroelectric states are also connected by  ${\cal O}=\{{\cal C}_{2x}\mathcal{T}|\frac{2}{3}\frac{2}{3}\}$.
Charge and spin transport in bilayer CrI$_3$ are determined jointly by the point-group operation ${\cal C}_{2x}\mathcal{T}$  and  $\{E|\frac{2}{3}\frac{2}{3}\}$.
Since the polarizations of state I are approximately $(P_a,P_b)=(0,0)\pmod{{\bm Q}}$ and $(P_a^s,P_b^s)=(0,0)\pmod{{\bm Q}}$,  we have
\begin{align}
{\cal C}_{2x}\mathcal{T}{\bm P}-{\bm P}=(0,0), \ \ \ {\cal C}_{2x}\mathcal{T}{\bm P}^s-{\bm P}^s=(0,0),
\end{align}
indicating that the variation of polarizations must be $0$ or integer multiples of ${\bm Q}$, according to the definition of polarization in Eq. (\ref{EP}).
However, the presence of $\{E|\frac{2}{3}\frac{2}{3}0\}$ in ${\cal O}$ indicates a significant change in the spin electronic polarization of $n^s \times (\frac{2}{3}Q_a,\frac{2}{3}Q_b) \pmod{{\bm Q}}$, with $n^s=12$.
Thus,  the total variation $\delta {\bm P}^s$ obtained from the symmetry analysis is an integer multiple of ${\bm Q}$.

We then perform numerical calculations to verify the symmetry analysis, and unveil the underlying mechanism of pure spin transport.
In Figs.~\ref{fig:CrI3}(c-d), we plot the variations of ${\bm P}^{\rm ion}$, ${\bm P}^{\uparrow}$, ${\bm P}^{\downarrow}$, ${\bm P}$ and ${\bm P}^{\rm s}$ along the switching path, respectively.
All these variations contain solely the $(10)$ component.
Although all the polarizations exhibit sizable variations along the path, the total electric polarization along $(10)$ is nearly unchanged and returns to zero in the final state, i.e.  $\delta {\bm P}=(0,0)$ [see Fig.~\ref{fig:CrI3}(c)].
However, $\delta {\bm P}^s$ is nonzero and equals an integer multiple of ${\bm Q}$:  $\delta {\bm P}^s=(-2Q_a,0)$, as shown in Fig.~\ref{fig:CrI3}(d).

Figure~\ref{fig:CrI3}(c) also reveals that the pure sliding spin current arises from the compensation between the electric-polarization contributions of valence electrons and ions, which is distinct from the mechanism of the SHE \cite{sinova2015spinb}.
Specifically, during switching, the ionic and the electronic charge centers in the top CrI$_3$ layer move coherently. Consequently, their charge-polarization contributions compensate each other, as shown in Fig.~\ref{fig:CrI3}(c). In contrast, the spin electronic polarization cannot be canceled by the ions, as ions exhibit negligible magnetic moment.

\begin{table*}[t]
\caption{\label{tab:table2} Material candidates for the sliding spin current. Here, $\delta {\bm P}^{s}$ is given in units of the polarization quantum. $\delta {\bm r}$ denotes the interlayer sliding by shifting the top layer. $q$ denotes the net charge of the top layer. All positions are given in fractional coordinates.}
\footnotesize
\renewcommand{\arraystretch}{1.5}
\begin{ruledtabular}
\begin{tabular}{lccccc}
Materials
& Magnetism
& $\delta {\bm P}^{s}$
& $\delta {\bm r}$
& $ q/e $
& ${\cal O}$\\
\colrule

$H$-stacked bilayer CrI$_3$
& FM & $(-2,0)$ & $(-1/3,0)$ & $\simeq-0.05$
& $\{\mathcal{C}_{2x}\mathcal{T}|\frac23\frac23\}$ \\

Bilayer 2H-V$X_2$ ($X$ = S, Se, Te)
& cFiM & $(1/3,-1/3)$ & $(-1/3,1/3)$ & $\simeq -0.02/-0.02/-0.03$
& $\{{\cal M}_z {\cal T}|\frac13\bar{\frac13}\}$  \\

Bilayer VSi$_2$$X_4$ ($X$ = N, P)
& cFiM & $(1/3,-1/3)$ & $(-1/3,1/3)$ & $\simeq 0.02/0.06$
& $\{{\cal M}_z {\cal T}|\frac13\bar{\frac13}\}$ \\

Bilayer NiI$_2$
& cFiM & $(-2/3,2/3)$ & $(1/3,-1/3)$ & $\simeq -0.02$
& $\{{\cal M}_z {\cal T}|\bar{\frac13}\bar{\frac23}\}$  \\

Bilayer Cr$_2$Ge$_2$Te$_6$
& cFiM & $(-2,-2)$ & $(1/3,1/3)$ & $\simeq -0.07$
& $\{{\cal M}_z {\cal T}|\frac23\frac23\}$ \\

Bilayer ScI$_2$
& FM & $(1/3,-1/3)$ & $(1/3,-1/3)$ & $\simeq -0.01$
& $\{{\cal M}_z|\bar{\frac13}\bar{\frac23}\}$ \\

\end{tabular}
\end{ruledtabular}
\end{table*}

\begin{figure}[t]
\includegraphics[width=0.5\textwidth]{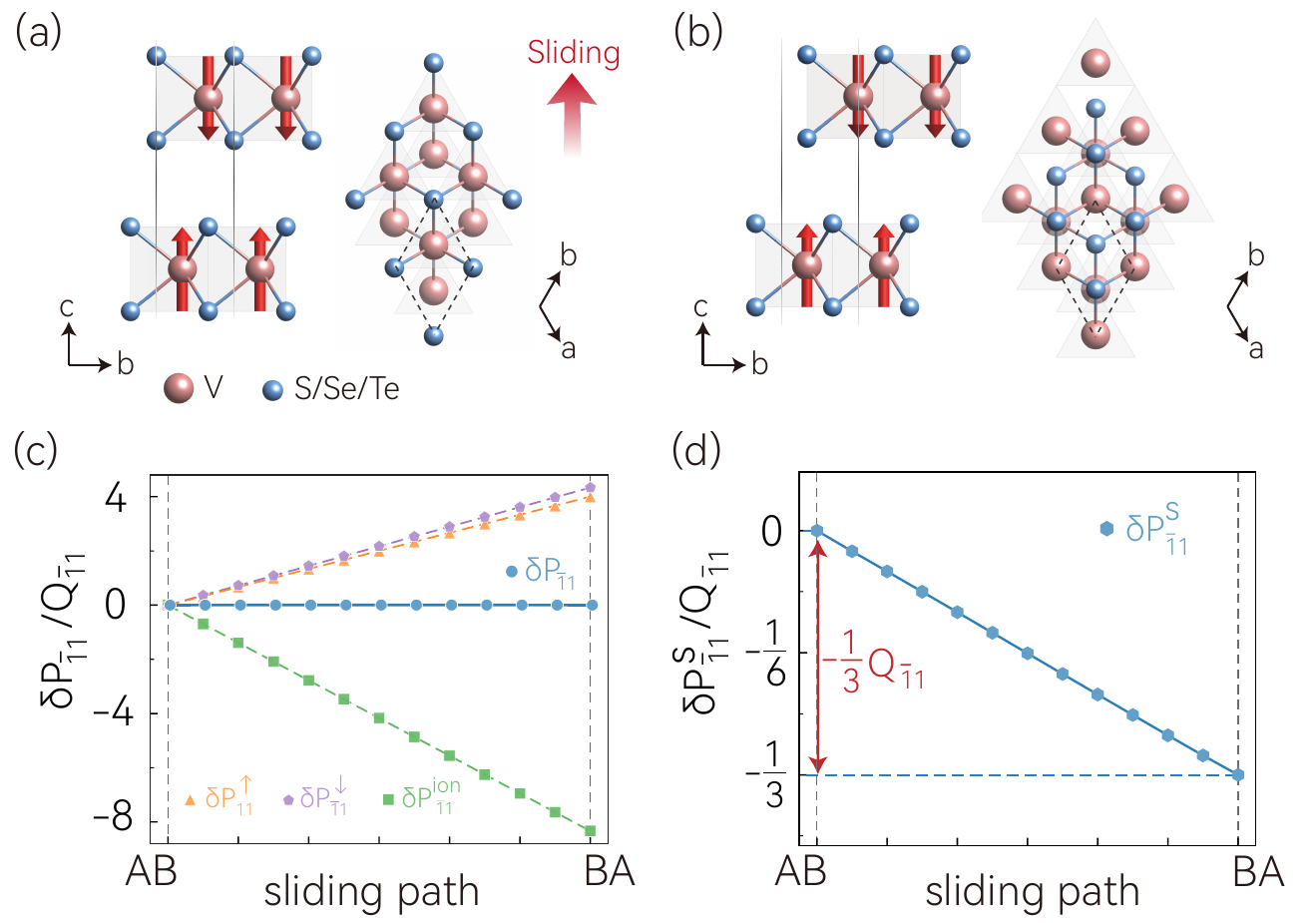}
\caption{\label{fig:VS2} (a-b)  Side and top views of bilayer $2H$-VS$_2$ in (a) AB stacking and (b) BA stacking.
The bilayer VS$_2$ can switch from AB stacking to BA stacking by shifting the top layer by $-\mathbf a/3+\mathbf b/3$, as marked in (a).
(c-d)  Variations of (c) electronic polarization and (d) spin electronic polarization along the ($\bar 1$1) direction.  $Q_a=1005.7$ pC/m denotes the polarization quantum along the ($\bar{1}1$) direction.}
\end{figure}

{\emph{\textcolor{blue}{$R$-stacked bilayer $2H$-V$X_2$.}--}}
We next study compensated ferrimagnetic (cFiM) bilayers of $2H$-V$X_2$ ($X=$ S, Se, Te), which have been predicted to be SFE candidates \cite{ma2024strain,sheng2025ubiquitous}.
Few-layer V$X_2$ have been experimentally synthesized \cite{zhang2017van,bonilla2018strong,liu2019observation,liu2019quasi2da,li2020structural,su2020submillimeterscale,zhu2022charge}.
Here, we take bilayer VS$_2$ as an example. The results for bilayer VSe$_2$ and VTe$_2$, which are similar to those for bilayer VS$_2$, are presented in the SM \cite{Hansupplemental}.
The ground state of  bilayer VS$_2$ is an interlayer antiferromagnetic state with compensated collinear magnetic order, as shown in Figs.~\ref{fig:VS2}(a-b).

Bilayer VS$_2$ belongs to magnetic space group No. 156.51 and preserves ${\cal C}_{3z}$ and ${\cal M}_{110}\mathcal{T}$ symmetries. The magnetic unit cell contains two V atoms and four S atoms.
The local moments are mainly localized on the V atoms, with an easy axis along the $z$ direction.
The V atom in the top (bottom) layer has a local moment of  $-1.2 \mu_B$  ($1.2 \mu_B$).
Since the two V atoms with opposite  magnetic moments occupy two different Wyckoff positions, they are not connected by symmetry, making the system a compensated ferrimagnet rather than an antiferromagnet \cite{vanleuken1995halfmetallic,kawamura2024compensateda,yuan2024nonrelativistic,liu2025twodimensionalb}.

We refer to the two energy-degenerate ferroelectric phases of the bilayer VS$_2$ as AB and BA stacking states, illustrated in Figs.~\ref{fig:VS2}(a) and \ref{fig:VS2}(b), respectively.
The AB and BA stacking state have opposite electric polarization $P_z=\pm$ 0.65 pC/m.
In the presence of a suitable electric field along the $z$-direction, the bilayer VS$_2$ can switch from the AB stacking to the BA stacking by shifting the top VS$_2$ layer by $-\mathbf a/3+\mathbf b/3$.
Interestingly, during sliding, the system can evolve from a cFiM into the recently proposed type-IV magnet \cite{bai2025anomalous,tian2026symmetry} in the intermediate state, which possesses $\{{\cal M}_z{\cal T}|\frac{1}{2}\frac{1}{2}0\}$ symmetry and consequently exhibits spin-degenerate electronic bands~\cite{Hansupplemental}.

With the bottom layer fixed, the AB- and BA-stacked bilayers are related  by ${\cal O}=\{{\cal M}_z {\cal T}|\frac13\bar{\frac13}0\}$.
Charge and  spin transport in VS$_2$ are determined jointly by the point-group operation ${\cal M}_z{\cal T}$ and  $\{E|\frac13\bar{\frac13}0\}$.
For  ${\cal M}_z {\cal T}$,  one has ${\cal M}_z{\cal T}P_{a(b)}=P_{a(b)}$ but ${\cal M}_z{\cal T}P_{a(b)}^s=-P_{a(b)}^s$.
In contrast,  $\{E|\frac13\bar{\frac13}0\}$ gives vanishing net charge and spin transport, as the system is both charge and spin neutral.
Thus, due to the presence of ${\cal C}_{3z}$ in both the AB- and BA-stacking, we have $\delta { P}_{a(b)}=0$ and $\delta {P}_{a(b)}^s=2{\bm F}_i \pmod{{\bm Q}}$ where ${\bm F}_1=(1/3,2/3)Q$ and ${\bm F}_2=(2/3,1/3)Q$  are the two ${\cal C}_{3z}$-allowed fractional polarization quanta \cite{pang2025generalized}.
This symmetry analysis directly suggests significant pure in-plane spin transport during ferroelectric switching.

In Figs.~\ref{fig:VS2}(c-d), we plot the variations of ${\bm P}^{\rm ion}$, ${\bm P}^{\uparrow}$, ${\bm P}^{\downarrow}$, ${\bm P}$ and ${\bm P}^{\rm s}$ along  the switching path, respectively.
All these variations contain solely the $(\bar{1}10)$ component.
One observes that while the total electric polarization $P_{\bar{1}10}$ remains unchanged [see Fig.~\ref{fig:VS2}(c)],  $\delta {\bm P}^s$ is a  fractional polarization quantum, $(1/3,-1/3)Q$, which is equal to $2{\bm F}_2$ mod $\bm Q$.

{\emph{\textcolor{blue}{Discussion.--}}}
We have proposed a new mechanism to realize pure spin current in magnetic SFEs.
Over the past few years, SFEs have been experimentally established in an expanding family of vdW materials~\cite{viznerstern2021interfacial,fei2018ferroelectricb,yasuda2021stackingengineeredc,
wang2022interfacial,wan2022roomtemperature,meng2022slidinga,sui2023sliding,yang2024ferroelectric,bian2024developinga,yasuda2024ultrafasta,bai2025subnanoseconda}, especially in the magnetic material $H$-stacked CrI$_3$ \cite{fox2025sliding}.
These experimental advances, together with the wide and growing family of magnetic SFEs, suggest that the sliding spin current proposed here should also be experimentally accessible in realistic vdW magnets, opening a new direction for low-dissipation, fully electrical spintronics. Furthermore, our calculations are not limited to bilayer CrI$_3$ and V$X_2$ ($X$ = S, Se, Te), but also demonstrate FSFE and the associated sliding spin current in several other magnetic SFEs, including cFiM bilayer VSi$_2X_4$ ($X=$ N, P), NiI$_2$, Cr$_2$Ge$_2$Te$_6$, and FM bilayer ScI$_2$, as detailed in the SM~\cite{Hansupplemental}.

In particular, the SFEs have been experimentally demonstrated to be fatigue resistant \cite{bian2024developinga,yasuda2024ultrafasta}, and electric-field switching in SFEs can reach sub-nanosecond timescales in experiments~\cite{yasuda2024ultrafasta,bai2025subnanoseconda}.
Taking $1~\mathrm{ns}$ as the switching time,  the spin-current density can reach $1.5\times10^9 (\hbar/2e)\mathrm{A/m^2}$ for $H$-stacked bilayer CrI$_3$ and $3.72\times10^8 (\hbar/2e)\mathrm{A/m^2}$ for bilayer VS$_2$.
Therefore, an alternating electric field can enable a fatigue-resistant, high-frequency alternating spin-current source with a high spin-current density in stacked CrI$_3$ and other magnetic SFEs.

Moreover, recent theoretical work shows that the light-induced ferroelectric switching in SFEs can enter the picosecond regime~\cite{yang2024lightinduced}.
If picosecond switching is achieved, the resulting spin current would be further enhanced to as high as $10^{12} (\hbar/2e)\mathrm{A/m^2}$  in $H$-stacked bilayer CrI$_3$.

\acknowledgments
{\emph{\textcolor{blue}{Acknowledgments.--}}}
This work is supported by National Key R\&D Program of China (Grant No. 2025YFA1411200),  the National Natural Science Foundation of China (Grants No. 125B2096, No. 12474040, No. 12504107, No. 12234003), and the Natural Science Foundation of Beijing (Grant No. 1252029).

\bibliography{mybib}

\begin{thebibliography}{89}%
\makeatletter
\providecommand \@ifxundefined [1]{%
 \@ifx{#1\undefined}
}%
\providecommand \@ifnum [1]{%
 \ifnum #1\expandafter \@firstoftwo
 \else \expandafter \@secondoftwo
 \fi
}%
\providecommand \@ifx [1]{%
 \ifx #1\expandafter \@firstoftwo
 \else \expandafter \@secondoftwo
 \fi
}%
\providecommand \natexlab [1]{#1}%
\providecommand \enquote  [1]{``#1''}%
\providecommand \bibnamefont  [1]{#1}%
\providecommand \bibfnamefont [1]{#1}%
\providecommand \citenamefont [1]{#1}%
\providecommand \href@noop [0]{\@secondoftwo}%
\providecommand \href [0]{\begingroup \@sanitize@url \@href}%
\providecommand \@href[1]{\@@startlink{#1}\@@href}%
\providecommand \@@href[1]{\endgroup#1\@@endlink}%
\providecommand \@sanitize@url [0]{\catcode `\\12\catcode `\$12\catcode
  `\&12\catcode `\#12\catcode `\^12\catcode `\_12\catcode `\%12\relax}%
\providecommand \@@startlink[1]{}%
\providecommand \@@endlink[0]{}%
\providecommand \url  [0]{\begingroup\@sanitize@url \@url }%
\providecommand \@url [1]{\endgroup\@href {#1}{\urlprefix }}%
\providecommand \urlprefix  [0]{URL }%
\providecommand \Eprint [0]{\href }%
\providecommand \doibase [0]{https://doi.org/}%
\providecommand \selectlanguage [0]{\@gobble}%
\providecommand \bibinfo  [0]{\@secondoftwo}%
\providecommand \bibfield  [0]{\@secondoftwo}%
\providecommand \translation [1]{[#1]}%
\providecommand \BibitemOpen [0]{}%
\providecommand \bibitemStop [0]{}%
\providecommand \bibitemNoStop [0]{.\EOS\space}%
\providecommand \EOS [0]{\spacefactor3000\relax}%
\providecommand \BibitemShut  [1]{\csname bibitem#1\endcsname}%
\let\auto@bib@innerbib\@empty
\bibitem [{\citenamefont {{\v Z}uti{\'c}}\ \emph {et~al.}(2004)\citenamefont
  {{\v Z}uti{\'c}}, \citenamefont {Fabian},\ and\ \citenamefont
  {Das~Sarma}}]{zutic2004spintronicsb}%
  \BibitemOpen
  \bibfield  {author} {\bibinfo {author} {\bibfnamefont {I.}~\bibnamefont {{\v
  Z}uti{\'c}}}, \bibinfo {author} {\bibfnamefont {J.}~\bibnamefont {Fabian}},\
  and\ \bibinfo {author} {\bibfnamefont {S.}~\bibnamefont {Das~Sarma}},\
  }\bibfield  {title} {\bibinfo {title} {Spintronics: {{Fundamentals}} and
  applications},\ }\href {https://doi.org/10.1103/RevModPhys.76.323} {\bibfield
   {journal} {\bibinfo  {journal} {Reviews of Modern Physics}\ }\textbf
  {\bibinfo {volume} {76}},\ \bibinfo {pages} {323} (\bibinfo {year}
  {2004})}\BibitemShut {NoStop}%
\bibitem [{\citenamefont {Fert}(2008)}]{fert2008nobel}%
  \BibitemOpen
  \bibfield  {author} {\bibinfo {author} {\bibfnamefont {A.}~\bibnamefont
  {Fert}},\ }\bibfield  {title} {\bibinfo {title} {Nobel {{Lecture}}:
  {{Origin}}, development, and future of spintronics},\ }\href
  {https://doi.org/10.1103/RevModPhys.80.1517} {\bibfield  {journal} {\bibinfo
  {journal} {Reviews of Modern Physics}\ }\textbf {\bibinfo {volume} {80}},\
  \bibinfo {pages} {1517} (\bibinfo {year} {2008})}\BibitemShut {NoStop}%
\bibitem [{\citenamefont {Bader}\ and\ \citenamefont
  {Parkin}(2010)}]{bader2010spintronicsb}%
  \BibitemOpen
  \bibfield  {author} {\bibinfo {author} {\bibfnamefont {S.~D.}\ \bibnamefont
  {Bader}}\ and\ \bibinfo {author} {\bibfnamefont {S.~S.~P.}\ \bibnamefont
  {Parkin}},\ }\bibfield  {title} {\bibinfo {title} {Spintronics},\ }\href
  {https://doi.org/10.1146/annurev-conmatphys-070909-104123} {\bibfield
  {journal} {\bibinfo  {journal} {Annual Review of Condensed Matter Physics}\
  }\textbf {\bibinfo {volume} {1}},\ \bibinfo {pages} {71} (\bibinfo {year}
  {2010})}\BibitemShut {NoStop}%
\bibitem [{\citenamefont {Hirohata}\ \emph {et~al.}(2020)\citenamefont
  {Hirohata}, \citenamefont {Yamada}, \citenamefont {Nakatani}, \citenamefont
  {Prejbeanu}, \citenamefont {Di{\'e}ny}, \citenamefont {Pirro},\ and\
  \citenamefont {Hillebrands}}]{hirohata2020reviewa}%
  \BibitemOpen
  \bibfield  {author} {\bibinfo {author} {\bibfnamefont {A.}~\bibnamefont
  {Hirohata}}, \bibinfo {author} {\bibfnamefont {K.}~\bibnamefont {Yamada}},
  \bibinfo {author} {\bibfnamefont {Y.}~\bibnamefont {Nakatani}}, \bibinfo
  {author} {\bibfnamefont {I.-L.}\ \bibnamefont {Prejbeanu}}, \bibinfo {author}
  {\bibfnamefont {B.}~\bibnamefont {Di{\'e}ny}}, \bibinfo {author}
  {\bibfnamefont {P.}~\bibnamefont {Pirro}},\ and\ \bibinfo {author}
  {\bibfnamefont {B.}~\bibnamefont {Hillebrands}},\ }\bibfield  {title}
  {\bibinfo {title} {Review on spintronics: {{Principles}} and device
  applications},\ }\href {https://doi.org/10.1016/j.jmmm.2020.166711}
  {\bibfield  {journal} {\bibinfo  {journal} {Journal of Magnetism and Magnetic
  Materials}\ }\textbf {\bibinfo {volume} {509}},\ \bibinfo {pages} {166711}
  (\bibinfo {year} {2020})}\BibitemShut {NoStop}%
\bibitem [{\citenamefont {Hirsch}(1999)}]{hirsch1999spin}%
  \BibitemOpen
  \bibfield  {author} {\bibinfo {author} {\bibfnamefont {J.~E.}\ \bibnamefont
  {Hirsch}},\ }\bibfield  {title} {\bibinfo {title} {Spin {{Hall Effect}}},\
  }\href {https://doi.org/10.1103/PhysRevLett.83.1834} {\bibfield  {journal}
  {\bibinfo  {journal} {Physical Review Letters}\ }\textbf {\bibinfo {volume}
  {83}},\ \bibinfo {pages} {1834} (\bibinfo {year} {1999})}\BibitemShut
  {NoStop}%
\bibitem [{\citenamefont {Tserkovnyak}\ \emph {et~al.}(2005)\citenamefont
  {Tserkovnyak}, \citenamefont {Brataas}, \citenamefont {Bauer},\ and\
  \citenamefont {Halperin}}]{tserkovnyak2005nonlocal}%
  \BibitemOpen
  \bibfield  {author} {\bibinfo {author} {\bibfnamefont {Y.}~\bibnamefont
  {Tserkovnyak}}, \bibinfo {author} {\bibfnamefont {A.}~\bibnamefont
  {Brataas}}, \bibinfo {author} {\bibfnamefont {G.~E.~W.}\ \bibnamefont
  {Bauer}},\ and\ \bibinfo {author} {\bibfnamefont {B.~I.}\ \bibnamefont
  {Halperin}},\ }\bibfield  {title} {\bibinfo {title} {Nonlocal magnetization
  dynamics in ferromagnetic heterostructures},\ }\href
  {https://doi.org/10.1103/RevModPhys.77.1375} {\bibfield  {journal} {\bibinfo
  {journal} {Reviews of Modern Physics}\ }\textbf {\bibinfo {volume} {77}},\
  \bibinfo {pages} {1375} (\bibinfo {year} {2005})}\BibitemShut {NoStop}%
\bibitem [{\citenamefont {Lee}\ \emph {et~al.}(2021)\citenamefont {Lee},
  \citenamefont {Kang}, \citenamefont {Go}, \citenamefont {Kim}, \citenamefont
  {Kang}, \citenamefont {Lee}, \citenamefont {Lee}, \citenamefont {Kang},
  \citenamefont {Lee}, \citenamefont {Mokrousov}, \citenamefont {Kim},
  \citenamefont {Kim}, \citenamefont {Lee},\ and\ \citenamefont
  {Park}}]{lee2021efficient}%
  \BibitemOpen
  \bibfield  {author} {\bibinfo {author} {\bibfnamefont {S.}~\bibnamefont
  {Lee}}, \bibinfo {author} {\bibfnamefont {M.-G.}\ \bibnamefont {Kang}},
  \bibinfo {author} {\bibfnamefont {D.}~\bibnamefont {Go}}, \bibinfo {author}
  {\bibfnamefont {D.}~\bibnamefont {Kim}}, \bibinfo {author} {\bibfnamefont
  {J.-H.}\ \bibnamefont {Kang}}, \bibinfo {author} {\bibfnamefont
  {T.}~\bibnamefont {Lee}}, \bibinfo {author} {\bibfnamefont {G.-H.}\
  \bibnamefont {Lee}}, \bibinfo {author} {\bibfnamefont {J.}~\bibnamefont
  {Kang}}, \bibinfo {author} {\bibfnamefont {N.~J.}\ \bibnamefont {Lee}},
  \bibinfo {author} {\bibfnamefont {Y.}~\bibnamefont {Mokrousov}}, \bibinfo
  {author} {\bibfnamefont {S.}~\bibnamefont {Kim}}, \bibinfo {author}
  {\bibfnamefont {K.-J.}\ \bibnamefont {Kim}}, \bibinfo {author} {\bibfnamefont
  {K.-J.}\ \bibnamefont {Lee}},\ and\ \bibinfo {author} {\bibfnamefont {B.-G.}\
  \bibnamefont {Park}},\ }\bibfield  {title} {\bibinfo {title} {Efficient
  conversion of orbital {{Hall}} current to spin current for spin-orbit torque
  switching},\ }\href {https://doi.org/10.1038/s42005-021-00737-7} {\bibfield
  {journal} {\bibinfo  {journal} {Communications Physics}\ }\textbf {\bibinfo
  {volume} {4}},\ \bibinfo {pages} {234} (\bibinfo {year} {2021})}\BibitemShut
  {NoStop}%
\bibitem [{\citenamefont {Bose}\ \emph {et~al.}(2022)\citenamefont {Bose},
  \citenamefont {Schreiber}, \citenamefont {Jain}, \citenamefont {Shao},
  \citenamefont {Nair}, \citenamefont {Sun}, \citenamefont {Zhang},
  \citenamefont {Muller}, \citenamefont {Tsymbal}, \citenamefont {Schlom},\
  and\ \citenamefont {Ralph}}]{bose2022tilted}%
  \BibitemOpen
  \bibfield  {author} {\bibinfo {author} {\bibfnamefont {A.}~\bibnamefont
  {Bose}}, \bibinfo {author} {\bibfnamefont {N.~J.}\ \bibnamefont {Schreiber}},
  \bibinfo {author} {\bibfnamefont {R.}~\bibnamefont {Jain}}, \bibinfo {author}
  {\bibfnamefont {D.-F.}\ \bibnamefont {Shao}}, \bibinfo {author}
  {\bibfnamefont {H.~P.}\ \bibnamefont {Nair}}, \bibinfo {author}
  {\bibfnamefont {J.}~\bibnamefont {Sun}}, \bibinfo {author} {\bibfnamefont
  {X.~S.}\ \bibnamefont {Zhang}}, \bibinfo {author} {\bibfnamefont {D.~A.}\
  \bibnamefont {Muller}}, \bibinfo {author} {\bibfnamefont {E.~Y.}\
  \bibnamefont {Tsymbal}}, \bibinfo {author} {\bibfnamefont {D.~G.}\
  \bibnamefont {Schlom}},\ and\ \bibinfo {author} {\bibfnamefont {D.~C.}\
  \bibnamefont {Ralph}},\ }\bibfield  {title} {\bibinfo {title} {Tilted spin
  current generated by the collinear antiferromagnet ruthenium dioxide},\
  }\href {https://doi.org/10.1038/s41928-022-00744-8} {\bibfield  {journal}
  {\bibinfo  {journal} {Nature Electronics}\ }\textbf {\bibinfo {volume} {5}},\
  \bibinfo {pages} {267} (\bibinfo {year} {2022})}\BibitemShut {NoStop}%
\bibitem [{\citenamefont {Kikkawa}\ and\ \citenamefont
  {Saitoh}(2023)}]{kikkawa2023spin}%
  \BibitemOpen
  \bibfield  {author} {\bibinfo {author} {\bibfnamefont {T.}~\bibnamefont
  {Kikkawa}}\ and\ \bibinfo {author} {\bibfnamefont {E.}~\bibnamefont
  {Saitoh}},\ }\bibfield  {title} {\bibinfo {title} {Spin {{Seebeck Effect}}:
  {{Sensitive Probe}} for {{Elementary Excitation}}, {{Spin Correlation}},
  {{Transport}}, {{Magnetic Order}}, and {{Domains}} in {{Solids}}},\ }\href
  {https://doi.org/10.1146/annurev-conmatphys-040721-014957} {\bibfield
  {journal} {\bibinfo  {journal} {Annual Review of Condensed Matter Physics}\
  }\textbf {\bibinfo {volume} {14}},\ \bibinfo {pages} {129} (\bibinfo {year}
  {2023})}\BibitemShut {NoStop}%
\bibitem [{\citenamefont {Ding}\ \emph {et~al.}(2024)\citenamefont {Ding},
  \citenamefont {Kang}, \citenamefont {Legrand},\ and\ \citenamefont
  {Gambardella}}]{ding2024orbital}%
  \BibitemOpen
  \bibfield  {author} {\bibinfo {author} {\bibfnamefont {S.}~\bibnamefont
  {Ding}}, \bibinfo {author} {\bibfnamefont {M.-G.}\ \bibnamefont {Kang}},
  \bibinfo {author} {\bibfnamefont {W.}~\bibnamefont {Legrand}},\ and\ \bibinfo
  {author} {\bibfnamefont {P.}~\bibnamefont {Gambardella}},\ }\bibfield
  {title} {\bibinfo {title} {Orbital {{Torque}} in {{Rare-Earth
  Transition-Metal Ferrimagnets}}},\ }\href
  {https://doi.org/10.1103/PhysRevLett.132.236702} {\bibfield  {journal}
  {\bibinfo  {journal} {Physical Review Letters}\ }\textbf {\bibinfo {volume}
  {132}},\ \bibinfo {pages} {236702} (\bibinfo {year} {2024})}\BibitemShut
  {NoStop}%
\bibitem [{\citenamefont {Chen}\ \emph {et~al.}(2025)\citenamefont {Chen},
  \citenamefont {Ji}, \citenamefont {Hong}, \citenamefont {Wan},\ and\
  \citenamefont {Xiang}}]{chen2025generation}%
  \BibitemOpen
  \bibfield  {author} {\bibinfo {author} {\bibfnamefont {Y.}~\bibnamefont
  {Chen}}, \bibinfo {author} {\bibfnamefont {J.}~\bibnamefont {Ji}}, \bibinfo
  {author} {\bibfnamefont {L.}~\bibnamefont {Hong}}, \bibinfo {author}
  {\bibfnamefont {X.}~\bibnamefont {Wan}},\ and\ \bibinfo {author}
  {\bibfnamefont {H.}~\bibnamefont {Xiang}},\ }\bibfield  {title} {\bibinfo
  {title} {Generation of {{Pure Spin Current}} with {{Insulating
  Antiferromagnetic Materials}}},\ }\href {https://doi.org/10.1103/n8d2-hjnd}
  {\bibfield  {journal} {\bibinfo  {journal} {Physical Review Letters}\
  }\textbf {\bibinfo {volume} {135}},\ \bibinfo {pages} {146703} (\bibinfo
  {year} {2025})}\BibitemShut {NoStop}%
\bibitem [{\citenamefont {Zhang}\ \emph
  {et~al.}(2025{\natexlab{a}})\citenamefont {Zhang}, \citenamefont {Bai},
  \citenamefont {Dai}, \citenamefont {Han}, \citenamefont {Chen}, \citenamefont
  {Liang}, \citenamefont {Cao}, \citenamefont {Zhang}, \citenamefont {Wang},
  \citenamefont {Zhu}, \citenamefont {Pan},\ and\ \citenamefont
  {Song}}]{zhang2025electrical}%
  \BibitemOpen
  \bibfield  {author} {\bibinfo {author} {\bibfnamefont {Y.}~\bibnamefont
  {Zhang}}, \bibinfo {author} {\bibfnamefont {H.}~\bibnamefont {Bai}}, \bibinfo
  {author} {\bibfnamefont {J.}~\bibnamefont {Dai}}, \bibinfo {author}
  {\bibfnamefont {L.}~\bibnamefont {Han}}, \bibinfo {author} {\bibfnamefont
  {C.}~\bibnamefont {Chen}}, \bibinfo {author} {\bibfnamefont {S.}~\bibnamefont
  {Liang}}, \bibinfo {author} {\bibfnamefont {Y.}~\bibnamefont {Cao}}, \bibinfo
  {author} {\bibfnamefont {Y.}~\bibnamefont {Zhang}}, \bibinfo {author}
  {\bibfnamefont {Q.}~\bibnamefont {Wang}}, \bibinfo {author} {\bibfnamefont
  {W.}~\bibnamefont {Zhu}}, \bibinfo {author} {\bibfnamefont {F.}~\bibnamefont
  {Pan}},\ and\ \bibinfo {author} {\bibfnamefont {C.}~\bibnamefont {Song}},\
  }\bibfield  {title} {\bibinfo {title} {Electrical manipulation of spin
  splitting torque in altermagnetic {{RuO2}}},\ }\href
  {https://doi.org/10.1038/s41467-025-60891-2} {\bibfield  {journal} {\bibinfo
  {journal} {Nature Communications}\ }\textbf {\bibinfo {volume} {16}},\
  \bibinfo {pages} {5646} (\bibinfo {year} {2025}{\natexlab{a}})}\BibitemShut
  {NoStop}%
\bibitem [{\citenamefont {Wang}\ \emph
  {et~al.}(2025{\natexlab{a}})\citenamefont {Wang}, \citenamefont {Liu},
  \citenamefont {Feng}, \citenamefont {Cao}, \citenamefont {Wu}, \citenamefont
  {Lai}, \citenamefont {Gao}, \citenamefont {Xiao},\ and\ \citenamefont
  {Yang}}]{wang2025intrinsic}%
  \BibitemOpen
  \bibfield  {author} {\bibinfo {author} {\bibfnamefont {H.}~\bibnamefont
  {Wang}}, \bibinfo {author} {\bibfnamefont {H.}~\bibnamefont {Liu}}, \bibinfo
  {author} {\bibfnamefont {X.}~\bibnamefont {Feng}}, \bibinfo {author}
  {\bibfnamefont {J.}~\bibnamefont {Cao}}, \bibinfo {author} {\bibfnamefont
  {W.}~\bibnamefont {Wu}}, \bibinfo {author} {\bibfnamefont {S.}~\bibnamefont
  {Lai}}, \bibinfo {author} {\bibfnamefont {W.}~\bibnamefont {Gao}}, \bibinfo
  {author} {\bibfnamefont {C.}~\bibnamefont {Xiao}},\ and\ \bibinfo {author}
  {\bibfnamefont {S.~A.}\ \bibnamefont {Yang}},\ }\bibfield  {title} {\bibinfo
  {title} {Intrinsic {{Nonlinear Spin Hall Effect}} and {{Manipulation}} of
  {{Perpendicular Magnetization}}},\ }\href
  {https://doi.org/10.1103/PhysRevLett.134.056301} {\bibfield  {journal}
  {\bibinfo  {journal} {Physical Review Letters}\ }\textbf {\bibinfo {volume}
  {134}},\ \bibinfo {pages} {056301} (\bibinfo {year}
  {2025}{\natexlab{a}})}\BibitemShut {NoStop}%
\bibitem [{\citenamefont {Sinova}\ \emph {et~al.}(2015)\citenamefont {Sinova},
  \citenamefont {Valenzuela}, \citenamefont {Wunderlich}, \citenamefont
  {Back},\ and\ \citenamefont {Jungwirth}}]{sinova2015spinb}%
  \BibitemOpen
  \bibfield  {author} {\bibinfo {author} {\bibfnamefont {J.}~\bibnamefont
  {Sinova}}, \bibinfo {author} {\bibfnamefont {S.~O.}\ \bibnamefont
  {Valenzuela}}, \bibinfo {author} {\bibfnamefont {J.}~\bibnamefont
  {Wunderlich}}, \bibinfo {author} {\bibfnamefont {C.~H.}\ \bibnamefont
  {Back}},\ and\ \bibinfo {author} {\bibfnamefont {T.}~\bibnamefont
  {Jungwirth}},\ }\bibfield  {title} {\bibinfo {title} {Spin {{Hall}}
  effects},\ }\href {https://doi.org/10.1103/RevModPhys.87.1213} {\bibfield
  {journal} {\bibinfo  {journal} {Reviews of Modern Physics}\ }\textbf
  {\bibinfo {volume} {87}},\ \bibinfo {pages} {1213} (\bibinfo {year}
  {2015})}\BibitemShut {NoStop}%
\bibitem [{\citenamefont {Kato}\ \emph {et~al.}(2004)\citenamefont {Kato},
  \citenamefont {Myers}, \citenamefont {Gossard},\ and\ \citenamefont
  {Awschalom}}]{kato2004observation}%
  \BibitemOpen
  \bibfield  {author} {\bibinfo {author} {\bibfnamefont {Y.~K.}\ \bibnamefont
  {Kato}}, \bibinfo {author} {\bibfnamefont {R.~C.}\ \bibnamefont {Myers}},
  \bibinfo {author} {\bibfnamefont {A.~C.}\ \bibnamefont {Gossard}},\ and\
  \bibinfo {author} {\bibfnamefont {D.~D.}\ \bibnamefont {Awschalom}},\
  }\bibfield  {title} {\bibinfo {title} {Observation of the {{Spin Hall
  Effect}} in {{Semiconductors}}},\ }\href
  {https://doi.org/10.1126/science.1105514} {\bibfield  {journal} {\bibinfo
  {journal} {Science}\ }\textbf {\bibinfo {volume} {306}},\ \bibinfo {pages}
  {1910} (\bibinfo {year} {2004})}\BibitemShut {NoStop}%
\bibitem [{\citenamefont {Valenzuela}\ and\ \citenamefont
  {Tinkham}(2006)}]{valenzuela2006direct}%
  \BibitemOpen
  \bibfield  {author} {\bibinfo {author} {\bibfnamefont {S.~O.}\ \bibnamefont
  {Valenzuela}}\ and\ \bibinfo {author} {\bibfnamefont {M.}~\bibnamefont
  {Tinkham}},\ }\bibfield  {title} {\bibinfo {title} {Direct electronic
  measurement of the spin {{Hall}} effect},\ }\href
  {https://doi.org/10.1038/nature04937} {\bibfield  {journal} {\bibinfo
  {journal} {Nature}\ }\textbf {\bibinfo {volume} {442}},\ \bibinfo {pages}
  {176} (\bibinfo {year} {2006})}\BibitemShut {NoStop}%
\bibitem [{\citenamefont {Jungwirth}\ \emph {et~al.}(2012)\citenamefont
  {Jungwirth}, \citenamefont {Wunderlich},\ and\ \citenamefont
  {Olejn{\'i}k}}]{jungwirth2012spin}%
  \BibitemOpen
  \bibfield  {author} {\bibinfo {author} {\bibfnamefont {T.}~\bibnamefont
  {Jungwirth}}, \bibinfo {author} {\bibfnamefont {J.}~\bibnamefont
  {Wunderlich}},\ and\ \bibinfo {author} {\bibfnamefont {K.}~\bibnamefont
  {Olejn{\'i}k}},\ }\bibfield  {title} {\bibinfo {title} {Spin {{Hall}} effect
  devices},\ }\href {https://doi.org/10.1038/nmat3279} {\bibfield  {journal}
  {\bibinfo  {journal} {Nature Materials}\ }\textbf {\bibinfo {volume} {11}},\
  \bibinfo {pages} {382} (\bibinfo {year} {2012})}\BibitemShut {NoStop}%
\bibitem [{\citenamefont {Liu}\ \emph {et~al.}(2012)\citenamefont {Liu},
  \citenamefont {Pai}, \citenamefont {Li}, \citenamefont {Tseng}, \citenamefont
  {Ralph},\ and\ \citenamefont {Buhrman}}]{liu2012spintorque}%
  \BibitemOpen
  \bibfield  {author} {\bibinfo {author} {\bibfnamefont {L.}~\bibnamefont
  {Liu}}, \bibinfo {author} {\bibfnamefont {C.-F.}\ \bibnamefont {Pai}},
  \bibinfo {author} {\bibfnamefont {Y.}~\bibnamefont {Li}}, \bibinfo {author}
  {\bibfnamefont {H.~W.}\ \bibnamefont {Tseng}}, \bibinfo {author}
  {\bibfnamefont {D.~C.}\ \bibnamefont {Ralph}},\ and\ \bibinfo {author}
  {\bibfnamefont {R.~A.}\ \bibnamefont {Buhrman}},\ }\bibfield  {title}
  {\bibinfo {title} {Spin-{{Torque Switching}} with the {{Giant Spin Hall
  Effect}} of {{Tantalum}}},\ }\href {https://doi.org/10.1126/science.1218197}
  {\bibfield  {journal} {\bibinfo  {journal} {Science}\ }\textbf {\bibinfo
  {volume} {336}},\ \bibinfo {pages} {555} (\bibinfo {year}
  {2012})}\BibitemShut {NoStop}%
\bibitem [{\citenamefont {Zhu}\ \emph {et~al.}(2019)\citenamefont {Zhu},
  \citenamefont {Zhu}, \citenamefont {Sui}, \citenamefont {Ralph},\ and\
  \citenamefont {Buhrman}}]{zhu2019variationa}%
  \BibitemOpen
  \bibfield  {author} {\bibinfo {author} {\bibfnamefont {L.}~\bibnamefont
  {Zhu}}, \bibinfo {author} {\bibfnamefont {L.}~\bibnamefont {Zhu}}, \bibinfo
  {author} {\bibfnamefont {M.}~\bibnamefont {Sui}}, \bibinfo {author}
  {\bibfnamefont {D.~C.}\ \bibnamefont {Ralph}},\ and\ \bibinfo {author}
  {\bibfnamefont {R.~A.}\ \bibnamefont {Buhrman}},\ }\bibfield  {title}
  {\bibinfo {title} {Variation of the giant intrinsic spin {{Hall}}
  conductivity of {{Pt}} with carrier lifetime},\ }\href
  {https://doi.org/10.1126/sciadv.aav8025} {\bibfield  {journal} {\bibinfo
  {journal} {Science Advances}\ }\textbf {\bibinfo {volume} {5}},\ \bibinfo
  {pages} {eaav8025} (\bibinfo {year} {2019})}\BibitemShut {NoStop}%
\bibitem [{\citenamefont {Abdelwahab}\ \emph {et~al.}(2024)\citenamefont
  {Abdelwahab}, \citenamefont {Kumar}, \citenamefont {Bian}, \citenamefont
  {Zheng}, \citenamefont {Gao}, \citenamefont {Hu}, \citenamefont {McClelland},
  \citenamefont {Leng}, \citenamefont {Wilson}, \citenamefont {Yin},
  \citenamefont {Yang},\ and\ \citenamefont
  {Loh}}]{abdelwahab2024twodimensional}%
  \BibitemOpen
  \bibfield  {author} {\bibinfo {author} {\bibfnamefont {I.}~\bibnamefont
  {Abdelwahab}}, \bibinfo {author} {\bibfnamefont {D.}~\bibnamefont {Kumar}},
  \bibinfo {author} {\bibfnamefont {T.}~\bibnamefont {Bian}}, \bibinfo {author}
  {\bibfnamefont {H.}~\bibnamefont {Zheng}}, \bibinfo {author} {\bibfnamefont
  {H.}~\bibnamefont {Gao}}, \bibinfo {author} {\bibfnamefont {F.}~\bibnamefont
  {Hu}}, \bibinfo {author} {\bibfnamefont {A.}~\bibnamefont {McClelland}},
  \bibinfo {author} {\bibfnamefont {K.}~\bibnamefont {Leng}}, \bibinfo {author}
  {\bibfnamefont {W.~L.}\ \bibnamefont {Wilson}}, \bibinfo {author}
  {\bibfnamefont {J.}~\bibnamefont {Yin}}, \bibinfo {author} {\bibfnamefont
  {H.}~\bibnamefont {Yang}},\ and\ \bibinfo {author} {\bibfnamefont {K.~P.}\
  \bibnamefont {Loh}},\ }\bibfield  {title} {\bibinfo {title} {Two-dimensional
  chiral perovskites with large spin {{Hall}} angle and collinear spin {{Hall}}
  conductivity},\ }\href {https://doi.org/10.1126/science.adq0967} {\bibfield
  {journal} {\bibinfo  {journal} {Science}\ }\textbf {\bibinfo {volume}
  {385}},\ \bibinfo {pages} {311} (\bibinfo {year} {2024})}\BibitemShut
  {NoStop}%
\bibitem [{\citenamefont {Wang}\ \emph
  {et~al.}(2025{\natexlab{b}})\citenamefont {Wang}, \citenamefont {Liu},
  \citenamefont {Feng}, \citenamefont {Cao}, \citenamefont {Wu}, \citenamefont
  {Lai}, \citenamefont {Gao}, \citenamefont {Xiao},\ and\ \citenamefont
  {Yang}}]{wang2025intrinsica}%
  \BibitemOpen
  \bibfield  {author} {\bibinfo {author} {\bibfnamefont {H.}~\bibnamefont
  {Wang}}, \bibinfo {author} {\bibfnamefont {H.}~\bibnamefont {Liu}}, \bibinfo
  {author} {\bibfnamefont {X.}~\bibnamefont {Feng}}, \bibinfo {author}
  {\bibfnamefont {J.}~\bibnamefont {Cao}}, \bibinfo {author} {\bibfnamefont
  {W.}~\bibnamefont {Wu}}, \bibinfo {author} {\bibfnamefont {S.}~\bibnamefont
  {Lai}}, \bibinfo {author} {\bibfnamefont {W.}~\bibnamefont {Gao}}, \bibinfo
  {author} {\bibfnamefont {C.}~\bibnamefont {Xiao}},\ and\ \bibinfo {author}
  {\bibfnamefont {S.~A.}\ \bibnamefont {Yang}},\ }\bibfield  {title} {\bibinfo
  {title} {Intrinsic {{Nonlinear Spin Hall Effect}} and {{Manipulation}} of
  {{Perpendicular Magnetization}}},\ }\href
  {https://doi.org/10.1103/PhysRevLett.134.056301} {\bibfield  {journal}
  {\bibinfo  {journal} {Physical Review Letters}\ }\textbf {\bibinfo {volume}
  {134}},\ \bibinfo {pages} {056301} (\bibinfo {year}
  {2025}{\natexlab{b}})}\BibitemShut {NoStop}%
\bibitem [{\citenamefont {Baltz}\ \emph {et~al.}(2018)\citenamefont {Baltz},
  \citenamefont {Manchon}, \citenamefont {Tsoi}, \citenamefont {Moriyama},
  \citenamefont {Ono},\ and\ \citenamefont
  {Tserkovnyak}}]{baltz2018antiferromagnetica}%
  \BibitemOpen
  \bibfield  {author} {\bibinfo {author} {\bibfnamefont {V.}~\bibnamefont
  {Baltz}}, \bibinfo {author} {\bibfnamefont {A.}~\bibnamefont {Manchon}},
  \bibinfo {author} {\bibfnamefont {M.}~\bibnamefont {Tsoi}}, \bibinfo {author}
  {\bibfnamefont {T.}~\bibnamefont {Moriyama}}, \bibinfo {author}
  {\bibfnamefont {T.}~\bibnamefont {Ono}},\ and\ \bibinfo {author}
  {\bibfnamefont {Y.}~\bibnamefont {Tserkovnyak}},\ }\bibfield  {title}
  {\bibinfo {title} {Antiferromagnetic spintronics},\ }\href
  {https://doi.org/10.1103/RevModPhys.90.015005} {\bibfield  {journal}
  {\bibinfo  {journal} {Reviews of Modern Physics}\ }\textbf {\bibinfo {volume}
  {90}},\ \bibinfo {pages} {015005} (\bibinfo {year} {2018})}\BibitemShut
  {NoStop}%
\bibitem [{\citenamefont {Tserkovnyak}\ \emph {et~al.}(2002)\citenamefont
  {Tserkovnyak}, \citenamefont {Brataas},\ and\ \citenamefont
  {Bauer}}]{tserkovnyak2002enhanced}%
  \BibitemOpen
  \bibfield  {author} {\bibinfo {author} {\bibfnamefont {Y.}~\bibnamefont
  {Tserkovnyak}}, \bibinfo {author} {\bibfnamefont {A.}~\bibnamefont
  {Brataas}},\ and\ \bibinfo {author} {\bibfnamefont {G.~E.~W.}\ \bibnamefont
  {Bauer}},\ }\bibfield  {title} {\bibinfo {title} {Enhanced {{Gilbert
  Damping}} in {{Thin Ferromagnetic Films}}},\ }\href
  {https://doi.org/10.1103/PhysRevLett.88.117601} {\bibfield  {journal}
  {\bibinfo  {journal} {Physical Review Letters}\ }\textbf {\bibinfo {volume}
  {88}},\ \bibinfo {pages} {117601} (\bibinfo {year} {2002})}\BibitemShut
  {NoStop}%
\bibitem [{\citenamefont {Mosendz}\ \emph {et~al.}(2010)\citenamefont
  {Mosendz}, \citenamefont {Pearson}, \citenamefont {Fradin}, \citenamefont
  {Bauer}, \citenamefont {Bader},\ and\ \citenamefont
  {Hoffmann}}]{mosendz2010quantifying}%
  \BibitemOpen
  \bibfield  {author} {\bibinfo {author} {\bibfnamefont {O.}~\bibnamefont
  {Mosendz}}, \bibinfo {author} {\bibfnamefont {J.~E.}\ \bibnamefont
  {Pearson}}, \bibinfo {author} {\bibfnamefont {F.~Y.}\ \bibnamefont {Fradin}},
  \bibinfo {author} {\bibfnamefont {G.~E.~W.}\ \bibnamefont {Bauer}}, \bibinfo
  {author} {\bibfnamefont {S.~D.}\ \bibnamefont {Bader}},\ and\ \bibinfo
  {author} {\bibfnamefont {A.}~\bibnamefont {Hoffmann}},\ }\bibfield  {title}
  {\bibinfo {title} {Quantifying {{Spin Hall Angles}} from {{Spin Pumping}}:
  {{Experiments}} and {{Theory}}},\ }\href
  {https://doi.org/10.1103/PhysRevLett.104.046601} {\bibfield  {journal}
  {\bibinfo  {journal} {Physical Review Letters}\ }\textbf {\bibinfo {volume}
  {104}},\ \bibinfo {pages} {046601} (\bibinfo {year} {2010})}\BibitemShut
  {NoStop}%
\bibitem [{\citenamefont {Cheng}\ \emph {et~al.}(2014)\citenamefont {Cheng},
  \citenamefont {Xiao}, \citenamefont {Niu},\ and\ \citenamefont
  {Brataas}}]{cheng2014spin}%
  \BibitemOpen
  \bibfield  {author} {\bibinfo {author} {\bibfnamefont {R.}~\bibnamefont
  {Cheng}}, \bibinfo {author} {\bibfnamefont {J.}~\bibnamefont {Xiao}},
  \bibinfo {author} {\bibfnamefont {Q.}~\bibnamefont {Niu}},\ and\ \bibinfo
  {author} {\bibfnamefont {A.}~\bibnamefont {Brataas}},\ }\bibfield  {title}
  {\bibinfo {title} {Spin {{Pumping}} and {{Spin-Transfer Torques}} in
  {{Antiferromagnets}}},\ }\href
  {https://doi.org/10.1103/PhysRevLett.113.057601} {\bibfield  {journal}
  {\bibinfo  {journal} {Physical Review Letters}\ }\textbf {\bibinfo {volume}
  {113}},\ \bibinfo {pages} {057601} (\bibinfo {year} {2014})}\BibitemShut
  {NoStop}%
\bibitem [{\citenamefont {Hamara}\ \emph {et~al.}(2024)\citenamefont {Hamara},
  \citenamefont {Strungaru}, \citenamefont {Massey}, \citenamefont {Remy},
  \citenamefont {Chen}, \citenamefont {Nava~Antonio}, \citenamefont
  {Alves~Santos}, \citenamefont {Hehn}, \citenamefont {Evans}, \citenamefont
  {Chantrell}, \citenamefont {Mangin}, \citenamefont {Ducati}, \citenamefont
  {Marrows}, \citenamefont {Barker},\ and\ \citenamefont
  {Ciccarelli}}]{hamara2024ultrahigh}%
  \BibitemOpen
  \bibfield  {author} {\bibinfo {author} {\bibfnamefont {D.}~\bibnamefont
  {Hamara}}, \bibinfo {author} {\bibfnamefont {M.}~\bibnamefont {Strungaru}},
  \bibinfo {author} {\bibfnamefont {J.~R.}\ \bibnamefont {Massey}}, \bibinfo
  {author} {\bibfnamefont {Q.}~\bibnamefont {Remy}}, \bibinfo {author}
  {\bibfnamefont {X.}~\bibnamefont {Chen}}, \bibinfo {author} {\bibfnamefont
  {G.}~\bibnamefont {Nava~Antonio}}, \bibinfo {author} {\bibfnamefont
  {O.}~\bibnamefont {Alves~Santos}}, \bibinfo {author} {\bibfnamefont
  {M.}~\bibnamefont {Hehn}}, \bibinfo {author} {\bibfnamefont {R.~F.~L.}\
  \bibnamefont {Evans}}, \bibinfo {author} {\bibfnamefont {R.~W.}\ \bibnamefont
  {Chantrell}}, \bibinfo {author} {\bibfnamefont {S.}~\bibnamefont {Mangin}},
  \bibinfo {author} {\bibfnamefont {C.}~\bibnamefont {Ducati}}, \bibinfo
  {author} {\bibfnamefont {C.~H.}\ \bibnamefont {Marrows}}, \bibinfo {author}
  {\bibfnamefont {J.}~\bibnamefont {Barker}},\ and\ \bibinfo {author}
  {\bibfnamefont {C.}~\bibnamefont {Ciccarelli}},\ }\bibfield  {title}
  {\bibinfo {title} {Ultra-high spin emission from antiferromagnetic
  {{FeRh}}},\ }\href {https://doi.org/10.1038/s41467-024-48795-z} {\bibfield
  {journal} {\bibinfo  {journal} {Nature Communications}\ }\textbf {\bibinfo
  {volume} {15}},\ \bibinfo {pages} {4958} (\bibinfo {year}
  {2024})}\BibitemShut {NoStop}%
\bibitem [{\citenamefont {Tang}\ and\ \citenamefont
  {Bauer}(2024)}]{tang2024thermal}%
  \BibitemOpen
  \bibfield  {author} {\bibinfo {author} {\bibfnamefont {P.}~\bibnamefont
  {Tang}}\ and\ \bibinfo {author} {\bibfnamefont {G.~E.~W.}\ \bibnamefont
  {Bauer}},\ }\bibfield  {title} {\bibinfo {title} {Thermal and {{Coherent Spin
  Pumping}} by {{Noncollinear Antiferromagnets}}},\ }\href
  {https://doi.org/10.1103/PhysRevLett.133.036701} {\bibfield  {journal}
  {\bibinfo  {journal} {Physical Review Letters}\ }\textbf {\bibinfo {volume}
  {133}},\ \bibinfo {pages} {036701} (\bibinfo {year} {2024})}\BibitemShut
  {NoStop}%
\bibitem [{\citenamefont {Li}\ and\ \citenamefont {Wu}(2017)}]{li2017binaryd}%
  \BibitemOpen
  \bibfield  {author} {\bibinfo {author} {\bibfnamefont {L.}~\bibnamefont
  {Li}}\ and\ \bibinfo {author} {\bibfnamefont {M.}~\bibnamefont {Wu}},\
  }\bibfield  {title} {\bibinfo {title} {Binary {{Compound Bilayer}} and
  {{Multilayer}} with {{Vertical Polarizations}}: {{Two-Dimensional
  Ferroelectrics}}, {{Multiferroics}}, and {{Nanogenerators}}},\ }\href
  {https://doi.org/10.1021/acsnano.7b02756} {\bibfield  {journal} {\bibinfo
  {journal} {ACS Nano}\ }\textbf {\bibinfo {volume} {11}},\ \bibinfo {pages}
  {6382} (\bibinfo {year} {2017})}\BibitemShut {NoStop}%
\bibitem [{\citenamefont {Wu}\ and\ \citenamefont {Li}(2021)}]{wu2021slidinga}%
  \BibitemOpen
  \bibfield  {author} {\bibinfo {author} {\bibfnamefont {M.}~\bibnamefont
  {Wu}}\ and\ \bibinfo {author} {\bibfnamefont {J.}~\bibnamefont {Li}},\
  }\bibfield  {title} {\bibinfo {title} {Sliding ferroelectricity in {{2D}} van
  der {{Waals}} materials: {{Related}} physics and future opportunities},\
  }\href {https://doi.org/10.1073/pnas.2115703118} {\bibfield  {journal}
  {\bibinfo  {journal} {Proceedings of the National Academy of Sciences}\
  }\textbf {\bibinfo {volume} {118}},\ \bibinfo {pages} {e2115703118} (\bibinfo
  {year} {2021})}\BibitemShut {NoStop}%
\bibitem [{\citenamefont {Vizner~Stern}\ \emph {et~al.}(2021)\citenamefont
  {Vizner~Stern}, \citenamefont {Waschitz}, \citenamefont {Cao}, \citenamefont
  {Nevo}, \citenamefont {Watanabe}, \citenamefont {Taniguchi}, \citenamefont
  {Sela}, \citenamefont {Urbakh}, \citenamefont {Hod},\ and\ \citenamefont
  {Ben~Shalom}}]{viznerstern2021interfacial}%
  \BibitemOpen
  \bibfield  {author} {\bibinfo {author} {\bibfnamefont {M.}~\bibnamefont
  {Vizner~Stern}}, \bibinfo {author} {\bibfnamefont {Y.}~\bibnamefont
  {Waschitz}}, \bibinfo {author} {\bibfnamefont {W.}~\bibnamefont {Cao}},
  \bibinfo {author} {\bibfnamefont {I.}~\bibnamefont {Nevo}}, \bibinfo {author}
  {\bibfnamefont {K.}~\bibnamefont {Watanabe}}, \bibinfo {author}
  {\bibfnamefont {T.}~\bibnamefont {Taniguchi}}, \bibinfo {author}
  {\bibfnamefont {E.}~\bibnamefont {Sela}}, \bibinfo {author} {\bibfnamefont
  {M.}~\bibnamefont {Urbakh}}, \bibinfo {author} {\bibfnamefont
  {O.}~\bibnamefont {Hod}},\ and\ \bibinfo {author} {\bibfnamefont
  {M.}~\bibnamefont {Ben~Shalom}},\ }\bibfield  {title} {\bibinfo {title}
  {Interfacial ferroelectricity by van der {{Waals}} sliding},\ }\href
  {https://doi.org/10.1126/science.abe8177} {\bibfield  {journal} {\bibinfo
  {journal} {Science}\ }\textbf {\bibinfo {volume} {372}},\ \bibinfo {pages}
  {1462} (\bibinfo {year} {2021})}\BibitemShut {NoStop}%
\bibitem [{\citenamefont {Qi}\ \emph {et~al.}(2021)\citenamefont {Qi},
  \citenamefont {Ruan},\ and\ \citenamefont {Zeng}}]{qi2021review}%
  \BibitemOpen
  \bibfield  {author} {\bibinfo {author} {\bibfnamefont {L.}~\bibnamefont
  {Qi}}, \bibinfo {author} {\bibfnamefont {S.}~\bibnamefont {Ruan}},\ and\
  \bibinfo {author} {\bibfnamefont {Y.-J.}\ \bibnamefont {Zeng}},\ }\bibfield
  {title} {\bibinfo {title} {Review on {{Recent Developments}} in {{2D
  Ferroelectrics}}: {{Theories}} and {{Applications}}},\ }\href
  {https://doi.org/10.1002/adma.202005098} {\bibfield  {journal} {\bibinfo
  {journal} {Advanced Materials}\ }\textbf {\bibinfo {volume} {33}},\ \bibinfo
  {pages} {2005098} (\bibinfo {year} {2021})}\BibitemShut {NoStop}%
\bibitem [{\citenamefont {Vizner~Stern}\ \emph {et~al.}(2025)\citenamefont
  {Vizner~Stern}, \citenamefont {Salleh~Atri},\ and\ \citenamefont
  {Ben~Shalom}}]{viznerstern2025sliding}%
  \BibitemOpen
  \bibfield  {author} {\bibinfo {author} {\bibfnamefont {M.}~\bibnamefont
  {Vizner~Stern}}, \bibinfo {author} {\bibfnamefont {S.}~\bibnamefont
  {Salleh~Atri}},\ and\ \bibinfo {author} {\bibfnamefont {M.}~\bibnamefont
  {Ben~Shalom}},\ }\bibfield  {title} {\bibinfo {title} {Sliding van der
  {{Waals}} polytypes},\ }\href {https://doi.org/10.1038/s42254-024-00781-6}
  {\bibfield  {journal} {\bibinfo  {journal} {Nature Reviews Physics}\ }\textbf
  {\bibinfo {volume} {7}},\ \bibinfo {pages} {50} (\bibinfo {year}
  {2025})}\BibitemShut {NoStop}%
\bibitem [{\citenamefont {Zhang}\ \emph
  {et~al.}(2025{\natexlab{b}})\citenamefont {Zhang}, \citenamefont {Fan},
  \citenamefont {Wang}, \citenamefont {Wu}, \citenamefont {Li}, \citenamefont
  {Meng},\ and\ \citenamefont {Geng}}]{zhang2025emerging}%
  \BibitemOpen
  \bibfield  {author} {\bibinfo {author} {\bibfnamefont {Q.}~\bibnamefont
  {Zhang}}, \bibinfo {author} {\bibfnamefont {A.}~\bibnamefont {Fan}}, \bibinfo
  {author} {\bibfnamefont {Y.}~\bibnamefont {Wang}}, \bibinfo {author}
  {\bibfnamefont {F.}~\bibnamefont {Wu}}, \bibinfo {author} {\bibfnamefont
  {L.}~\bibnamefont {Li}}, \bibinfo {author} {\bibfnamefont {H.}~\bibnamefont
  {Meng}},\ and\ \bibinfo {author} {\bibfnamefont {D.}~\bibnamefont {Geng}},\
  }\bibfield  {title} {\bibinfo {title} {Emerging frontiers in two-dimensional
  sliding ferroelectrics},\ }\href {https://doi.org/10.1038/s41699-025-00600-1}
  {\bibfield  {journal} {\bibinfo  {journal} {npj 2D Materials and
  Applications}\ }\textbf {\bibinfo {volume} {9}},\ \bibinfo {pages} {76}
  (\bibinfo {year} {2025}{\natexlab{b}})}\BibitemShut {NoStop}%
\bibitem [{\citenamefont {Ji}\ \emph {et~al.}(2024)\citenamefont {Ji},
  \citenamefont {Yu}, \citenamefont {Xu},\ and\ \citenamefont
  {Xiang}}]{ji2024fractionala}%
  \BibitemOpen
  \bibfield  {author} {\bibinfo {author} {\bibfnamefont {J.}~\bibnamefont
  {Ji}}, \bibinfo {author} {\bibfnamefont {G.}~\bibnamefont {Yu}}, \bibinfo
  {author} {\bibfnamefont {C.}~\bibnamefont {Xu}},\ and\ \bibinfo {author}
  {\bibfnamefont {H.~J.}\ \bibnamefont {Xiang}},\ }\bibfield  {title} {\bibinfo
  {title} {Fractional quantum ferroelectricity},\ }\href
  {https://doi.org/10.1038/s41467-023-44453-y} {\bibfield  {journal} {\bibinfo
  {journal} {Nature Communications}\ }\textbf {\bibinfo {volume} {15}},\
  \bibinfo {pages} {135} (\bibinfo {year} {2024})}\BibitemShut {NoStop}%
\bibitem [{\citenamefont {Luo}\ \emph {et~al.}(2026)\citenamefont {Luo},
  \citenamefont {Deng}, \citenamefont {Xiang},\ and\ \citenamefont
  {Bellaiche}}]{luo2026unified}%
  \BibitemOpen
  \bibfield  {author} {\bibinfo {author} {\bibfnamefont {W.}~\bibnamefont
  {Luo}}, \bibinfo {author} {\bibfnamefont {S.}~\bibnamefont {Deng}}, \bibinfo
  {author} {\bibfnamefont {H.}~\bibnamefont {Xiang}},\ and\ \bibinfo {author}
  {\bibfnamefont {L.}~\bibnamefont {Bellaiche}},\ }\href
  {https://doi.org/10.48550/arXiv.2605.14328} {\bibinfo {title} {Unified
  definition of ferroelectricity}} (\bibinfo {year} {2026}),\ \bibinfo {note}
  {comment: 22 pages, 4 figures},\ \Eprint {https://arxiv.org/abs/2605.14328}
  {arXiv:2605.14328 [cond-mat.mtrl-sci]} \BibitemShut {NoStop}%
\bibitem [{\citenamefont {Fei}\ \emph {et~al.}(2018)\citenamefont {Fei},
  \citenamefont {Zhao}, \citenamefont {Palomaki}, \citenamefont {Sun},
  \citenamefont {Miller}, \citenamefont {Zhao}, \citenamefont {Yan},
  \citenamefont {Xu},\ and\ \citenamefont {Cobden}}]{fei2018ferroelectricb}%
  \BibitemOpen
  \bibfield  {author} {\bibinfo {author} {\bibfnamefont {Z.}~\bibnamefont
  {Fei}}, \bibinfo {author} {\bibfnamefont {W.}~\bibnamefont {Zhao}}, \bibinfo
  {author} {\bibfnamefont {T.~A.}\ \bibnamefont {Palomaki}}, \bibinfo {author}
  {\bibfnamefont {B.}~\bibnamefont {Sun}}, \bibinfo {author} {\bibfnamefont
  {M.~K.}\ \bibnamefont {Miller}}, \bibinfo {author} {\bibfnamefont
  {Z.}~\bibnamefont {Zhao}}, \bibinfo {author} {\bibfnamefont {J.}~\bibnamefont
  {Yan}}, \bibinfo {author} {\bibfnamefont {X.}~\bibnamefont {Xu}},\ and\
  \bibinfo {author} {\bibfnamefont {D.~H.}\ \bibnamefont {Cobden}},\ }\bibfield
   {title} {\bibinfo {title} {Ferroelectric switching of a two-dimensional
  metal},\ }\href {https://doi.org/10.1038/s41586-018-0336-3} {\bibfield
  {journal} {\bibinfo  {journal} {Nature}\ }\textbf {\bibinfo {volume} {560}},\
  \bibinfo {pages} {336} (\bibinfo {year} {2018})}\BibitemShut {NoStop}%
\bibitem [{\citenamefont {Yasuda}\ \emph {et~al.}(2021)\citenamefont {Yasuda},
  \citenamefont {Wang}, \citenamefont {Watanabe}, \citenamefont {Taniguchi},\
  and\ \citenamefont {{Jarillo-Herrero}}}]{yasuda2021stackingengineeredc}%
  \BibitemOpen
  \bibfield  {author} {\bibinfo {author} {\bibfnamefont {K.}~\bibnamefont
  {Yasuda}}, \bibinfo {author} {\bibfnamefont {X.}~\bibnamefont {Wang}},
  \bibinfo {author} {\bibfnamefont {K.}~\bibnamefont {Watanabe}}, \bibinfo
  {author} {\bibfnamefont {T.}~\bibnamefont {Taniguchi}},\ and\ \bibinfo
  {author} {\bibfnamefont {P.}~\bibnamefont {{Jarillo-Herrero}}},\ }\bibfield
  {title} {\bibinfo {title} {Stacking-engineered ferroelectricity in bilayer
  boron nitride},\ }\href {https://doi.org/10.1126/science.abd3230} {\bibfield
  {journal} {\bibinfo  {journal} {Science}\ }\textbf {\bibinfo {volume}
  {372}},\ \bibinfo {pages} {1458} (\bibinfo {year} {2021})}\BibitemShut
  {NoStop}%
\bibitem [{\citenamefont {Meng}\ \emph {et~al.}(2022)\citenamefont {Meng},
  \citenamefont {Wu}, \citenamefont {Bian}, \citenamefont {Pan}, \citenamefont
  {Dong}, \citenamefont {Zhao}, \citenamefont {Chen}, \citenamefont {Wu},
  \citenamefont {Sun}, \citenamefont {Fu}, \citenamefont {Liu}, \citenamefont
  {Shi}, \citenamefont {Zhang}, \citenamefont {Zhang}, \citenamefont {Liu},\
  and\ \citenamefont {Liu}}]{meng2022slidinga}%
  \BibitemOpen
  \bibfield  {author} {\bibinfo {author} {\bibfnamefont {P.}~\bibnamefont
  {Meng}}, \bibinfo {author} {\bibfnamefont {Y.}~\bibnamefont {Wu}}, \bibinfo
  {author} {\bibfnamefont {R.}~\bibnamefont {Bian}}, \bibinfo {author}
  {\bibfnamefont {E.}~\bibnamefont {Pan}}, \bibinfo {author} {\bibfnamefont
  {B.}~\bibnamefont {Dong}}, \bibinfo {author} {\bibfnamefont {X.}~\bibnamefont
  {Zhao}}, \bibinfo {author} {\bibfnamefont {J.}~\bibnamefont {Chen}}, \bibinfo
  {author} {\bibfnamefont {L.}~\bibnamefont {Wu}}, \bibinfo {author}
  {\bibfnamefont {Y.}~\bibnamefont {Sun}}, \bibinfo {author} {\bibfnamefont
  {Q.}~\bibnamefont {Fu}}, \bibinfo {author} {\bibfnamefont {Q.}~\bibnamefont
  {Liu}}, \bibinfo {author} {\bibfnamefont {D.}~\bibnamefont {Shi}}, \bibinfo
  {author} {\bibfnamefont {Q.}~\bibnamefont {Zhang}}, \bibinfo {author}
  {\bibfnamefont {Y.-W.}\ \bibnamefont {Zhang}}, \bibinfo {author}
  {\bibfnamefont {Z.}~\bibnamefont {Liu}},\ and\ \bibinfo {author}
  {\bibfnamefont {F.}~\bibnamefont {Liu}},\ }\bibfield  {title} {\bibinfo
  {title} {Sliding induced multiple polarization states in two-dimensional
  ferroelectrics},\ }\href {https://doi.org/10.1038/s41467-022-35339-6}
  {\bibfield  {journal} {\bibinfo  {journal} {Nature Communications}\ }\textbf
  {\bibinfo {volume} {13}},\ \bibinfo {pages} {7696} (\bibinfo {year}
  {2022})}\BibitemShut {NoStop}%
\bibitem [{\citenamefont {Wan}\ \emph {et~al.}(2022)\citenamefont {Wan},
  \citenamefont {Hu}, \citenamefont {Mao}, \citenamefont {Fu}, \citenamefont
  {Yuan}, \citenamefont {Song}, \citenamefont {Gan}, \citenamefont {Xu},
  \citenamefont {Xue}, \citenamefont {Cheng}, \citenamefont {Huang},
  \citenamefont {Yang}, \citenamefont {Dai}, \citenamefont {Zeng},\ and\
  \citenamefont {Kan}}]{wan2022roomtemperature}%
  \BibitemOpen
  \bibfield  {author} {\bibinfo {author} {\bibfnamefont {Y.}~\bibnamefont
  {Wan}}, \bibinfo {author} {\bibfnamefont {T.}~\bibnamefont {Hu}}, \bibinfo
  {author} {\bibfnamefont {X.}~\bibnamefont {Mao}}, \bibinfo {author}
  {\bibfnamefont {J.}~\bibnamefont {Fu}}, \bibinfo {author} {\bibfnamefont
  {K.}~\bibnamefont {Yuan}}, \bibinfo {author} {\bibfnamefont {Y.}~\bibnamefont
  {Song}}, \bibinfo {author} {\bibfnamefont {X.}~\bibnamefont {Gan}}, \bibinfo
  {author} {\bibfnamefont {X.}~\bibnamefont {Xu}}, \bibinfo {author}
  {\bibfnamefont {M.}~\bibnamefont {Xue}}, \bibinfo {author} {\bibfnamefont
  {X.}~\bibnamefont {Cheng}}, \bibinfo {author} {\bibfnamefont
  {C.}~\bibnamefont {Huang}}, \bibinfo {author} {\bibfnamefont
  {J.}~\bibnamefont {Yang}}, \bibinfo {author} {\bibfnamefont {L.}~\bibnamefont
  {Dai}}, \bibinfo {author} {\bibfnamefont {H.}~\bibnamefont {Zeng}},\ and\
  \bibinfo {author} {\bibfnamefont {E.}~\bibnamefont {Kan}},\ }\bibfield
  {title} {\bibinfo {title} {Room-{{Temperature Ferroelectricity}} in
  {{$1T'$-ReS$_2$}} {{Multilayers}}},\ }\href
  {https://doi.org/10.1103/PhysRevLett.128.067601} {\bibfield  {journal}
  {\bibinfo  {journal} {Physical Review Letters}\ }\textbf {\bibinfo {volume}
  {128}},\ \bibinfo {pages} {067601} (\bibinfo {year} {2022})}\BibitemShut
  {NoStop}%
\bibitem [{\citenamefont {Wang}\ \emph {et~al.}(2022)\citenamefont {Wang},
  \citenamefont {Yasuda}, \citenamefont {Zhang}, \citenamefont {Liu},
  \citenamefont {Watanabe}, \citenamefont {Taniguchi}, \citenamefont {Hone},
  \citenamefont {Fu},\ and\ \citenamefont
  {{Jarillo-Herrero}}}]{wang2022interfacial}%
  \BibitemOpen
  \bibfield  {author} {\bibinfo {author} {\bibfnamefont {X.}~\bibnamefont
  {Wang}}, \bibinfo {author} {\bibfnamefont {K.}~\bibnamefont {Yasuda}},
  \bibinfo {author} {\bibfnamefont {Y.}~\bibnamefont {Zhang}}, \bibinfo
  {author} {\bibfnamefont {S.}~\bibnamefont {Liu}}, \bibinfo {author}
  {\bibfnamefont {K.}~\bibnamefont {Watanabe}}, \bibinfo {author}
  {\bibfnamefont {T.}~\bibnamefont {Taniguchi}}, \bibinfo {author}
  {\bibfnamefont {J.}~\bibnamefont {Hone}}, \bibinfo {author} {\bibfnamefont
  {L.}~\bibnamefont {Fu}},\ and\ \bibinfo {author} {\bibfnamefont
  {P.}~\bibnamefont {{Jarillo-Herrero}}},\ }\bibfield  {title} {\bibinfo
  {title} {Interfacial ferroelectricity in rhombohedral-stacked bilayer
  transition metal dichalcogenides},\ }\href
  {https://doi.org/10.1038/s41565-021-01059-z} {\bibfield  {journal} {\bibinfo
  {journal} {Nature Nanotechnology}\ }\textbf {\bibinfo {volume} {17}},\
  \bibinfo {pages} {367} (\bibinfo {year} {2022})}\BibitemShut {NoStop}%
\bibitem [{\citenamefont {Sui}\ \emph {et~al.}(2023)\citenamefont {Sui},
  \citenamefont {Jin}, \citenamefont {Zhang}, \citenamefont {Qi}, \citenamefont
  {Wu}, \citenamefont {Huang}, \citenamefont {Yue},\ and\ \citenamefont
  {Chu}}]{sui2023sliding}%
  \BibitemOpen
  \bibfield  {author} {\bibinfo {author} {\bibfnamefont {F.}~\bibnamefont
  {Sui}}, \bibinfo {author} {\bibfnamefont {M.}~\bibnamefont {Jin}}, \bibinfo
  {author} {\bibfnamefont {Y.}~\bibnamefont {Zhang}}, \bibinfo {author}
  {\bibfnamefont {R.}~\bibnamefont {Qi}}, \bibinfo {author} {\bibfnamefont
  {Y.-N.}\ \bibnamefont {Wu}}, \bibinfo {author} {\bibfnamefont
  {R.}~\bibnamefont {Huang}}, \bibinfo {author} {\bibfnamefont
  {F.}~\bibnamefont {Yue}},\ and\ \bibinfo {author} {\bibfnamefont
  {J.}~\bibnamefont {Chu}},\ }\bibfield  {title} {\bibinfo {title} {Sliding
  ferroelectricity in van der {{Waals}} layered {$\gamma$}-{{InSe}}
  semiconductor},\ }\href {https://doi.org/10.1038/s41467-022-35490-0}
  {\bibfield  {journal} {\bibinfo  {journal} {Nature Communications}\ }\textbf
  {\bibinfo {volume} {14}},\ \bibinfo {pages} {36} (\bibinfo {year}
  {2023})}\BibitemShut {NoStop}%
\bibitem [{\citenamefont {Yang}\ \emph {et~al.}(2024)\citenamefont {Yang},
  \citenamefont {Liang}, \citenamefont {Hu}, \citenamefont {Chen},
  \citenamefont {Ho}, \citenamefont {Chang}, \citenamefont {Yang},
  \citenamefont {Lo}, \citenamefont {Kuo}, \citenamefont {Chen}, \citenamefont
  {Lin}, \citenamefont {Simbulan}, \citenamefont {Luo}, \citenamefont {Chang},
  \citenamefont {Kuo}, \citenamefont {Ku}, \citenamefont {Chen}, \citenamefont
  {Huang}, \citenamefont {Chang}, \citenamefont {Chiang}, \citenamefont {Lu},
  \citenamefont {Lee}, \citenamefont {Li}, \citenamefont {Wu}, \citenamefont
  {Chen}, \citenamefont {Lin},\ and\ \citenamefont
  {Lan}}]{yang2024ferroelectric}%
  \BibitemOpen
  \bibfield  {author} {\bibinfo {author} {\bibfnamefont {T.~H.}\ \bibnamefont
  {Yang}}, \bibinfo {author} {\bibfnamefont {B.-W.}\ \bibnamefont {Liang}},
  \bibinfo {author} {\bibfnamefont {H.-C.}\ \bibnamefont {Hu}}, \bibinfo
  {author} {\bibfnamefont {F.-X.}\ \bibnamefont {Chen}}, \bibinfo {author}
  {\bibfnamefont {S.-Z.}\ \bibnamefont {Ho}}, \bibinfo {author} {\bibfnamefont
  {W.-H.}\ \bibnamefont {Chang}}, \bibinfo {author} {\bibfnamefont
  {L.}~\bibnamefont {Yang}}, \bibinfo {author} {\bibfnamefont {H.-C.}\
  \bibnamefont {Lo}}, \bibinfo {author} {\bibfnamefont {T.-H.}\ \bibnamefont
  {Kuo}}, \bibinfo {author} {\bibfnamefont {J.-H.}\ \bibnamefont {Chen}},
  \bibinfo {author} {\bibfnamefont {P.-Y.}\ \bibnamefont {Lin}}, \bibinfo
  {author} {\bibfnamefont {K.~B.}\ \bibnamefont {Simbulan}}, \bibinfo {author}
  {\bibfnamefont {Z.-F.}\ \bibnamefont {Luo}}, \bibinfo {author} {\bibfnamefont
  {A.~C.}\ \bibnamefont {Chang}}, \bibinfo {author} {\bibfnamefont {Y.-H.}\
  \bibnamefont {Kuo}}, \bibinfo {author} {\bibfnamefont {Y.-S.}\ \bibnamefont
  {Ku}}, \bibinfo {author} {\bibfnamefont {Y.-C.}\ \bibnamefont {Chen}},
  \bibinfo {author} {\bibfnamefont {Y.-J.}\ \bibnamefont {Huang}}, \bibinfo
  {author} {\bibfnamefont {Y.-C.}\ \bibnamefont {Chang}}, \bibinfo {author}
  {\bibfnamefont {Y.-F.}\ \bibnamefont {Chiang}}, \bibinfo {author}
  {\bibfnamefont {T.-H.}\ \bibnamefont {Lu}}, \bibinfo {author} {\bibfnamefont
  {M.-H.}\ \bibnamefont {Lee}}, \bibinfo {author} {\bibfnamefont {K.-S.}\
  \bibnamefont {Li}}, \bibinfo {author} {\bibfnamefont {M.}~\bibnamefont {Wu}},
  \bibinfo {author} {\bibfnamefont {Y.-C.}\ \bibnamefont {Chen}}, \bibinfo
  {author} {\bibfnamefont {C.-L.}\ \bibnamefont {Lin}},\ and\ \bibinfo {author}
  {\bibfnamefont {Y.-W.}\ \bibnamefont {Lan}},\ }\bibfield  {title} {\bibinfo
  {title} {Ferroelectric transistors based on shear-transformation-mediated
  rhombohedral-stacked molybdenum disulfide},\ }\href
  {https://doi.org/10.1038/s41928-023-01073-0} {\bibfield  {journal} {\bibinfo
  {journal} {Nature Electronics}\ }\textbf {\bibinfo {volume} {7}},\ \bibinfo
  {pages} {29} (\bibinfo {year} {2024})}\BibitemShut {NoStop}%
\bibitem [{\citenamefont {Bian}\ \emph {et~al.}(2024)\citenamefont {Bian},
  \citenamefont {He}, \citenamefont {Pan}, \citenamefont {Li}, \citenamefont
  {Cao}, \citenamefont {Meng}, \citenamefont {Chen}, \citenamefont {Liu},
  \citenamefont {Zhong}, \citenamefont {Li},\ and\ \citenamefont
  {Liu}}]{bian2024developinga}%
  \BibitemOpen
  \bibfield  {author} {\bibinfo {author} {\bibfnamefont {R.}~\bibnamefont
  {Bian}}, \bibinfo {author} {\bibfnamefont {R.}~\bibnamefont {He}}, \bibinfo
  {author} {\bibfnamefont {E.}~\bibnamefont {Pan}}, \bibinfo {author}
  {\bibfnamefont {Z.}~\bibnamefont {Li}}, \bibinfo {author} {\bibfnamefont
  {G.}~\bibnamefont {Cao}}, \bibinfo {author} {\bibfnamefont {P.}~\bibnamefont
  {Meng}}, \bibinfo {author} {\bibfnamefont {J.}~\bibnamefont {Chen}}, \bibinfo
  {author} {\bibfnamefont {Q.}~\bibnamefont {Liu}}, \bibinfo {author}
  {\bibfnamefont {Z.}~\bibnamefont {Zhong}}, \bibinfo {author} {\bibfnamefont
  {W.}~\bibnamefont {Li}},\ and\ \bibinfo {author} {\bibfnamefont
  {F.}~\bibnamefont {Liu}},\ }\bibfield  {title} {\bibinfo {title} {Developing
  fatigue-resistant ferroelectrics using interlayer sliding switching},\ }\href
  {https://doi.org/10.1126/science.ado1744} {\bibfield  {journal} {\bibinfo
  {journal} {Science}\ }\textbf {\bibinfo {volume} {385}},\ \bibinfo {pages}
  {57} (\bibinfo {year} {2024})}\BibitemShut {NoStop}%
\bibitem [{\citenamefont {Yasuda}\ \emph {et~al.}(2024)\citenamefont {Yasuda},
  \citenamefont {{Zalys-Geller}}, \citenamefont {Wang}, \citenamefont
  {Bennett}, \citenamefont {Cheema}, \citenamefont {Watanabe}, \citenamefont
  {Taniguchi}, \citenamefont {Kaxiras}, \citenamefont {{Jarillo-Herrero}},\
  and\ \citenamefont {Ashoori}}]{yasuda2024ultrafasta}%
  \BibitemOpen
  \bibfield  {author} {\bibinfo {author} {\bibfnamefont {K.}~\bibnamefont
  {Yasuda}}, \bibinfo {author} {\bibfnamefont {E.}~\bibnamefont
  {{Zalys-Geller}}}, \bibinfo {author} {\bibfnamefont {X.}~\bibnamefont
  {Wang}}, \bibinfo {author} {\bibfnamefont {D.}~\bibnamefont {Bennett}},
  \bibinfo {author} {\bibfnamefont {S.~S.}\ \bibnamefont {Cheema}}, \bibinfo
  {author} {\bibfnamefont {K.}~\bibnamefont {Watanabe}}, \bibinfo {author}
  {\bibfnamefont {T.}~\bibnamefont {Taniguchi}}, \bibinfo {author}
  {\bibfnamefont {E.}~\bibnamefont {Kaxiras}}, \bibinfo {author} {\bibfnamefont
  {P.}~\bibnamefont {{Jarillo-Herrero}}},\ and\ \bibinfo {author}
  {\bibfnamefont {R.}~\bibnamefont {Ashoori}},\ }\bibfield  {title} {\bibinfo
  {title} {Ultrafast high-endurance memory based on sliding ferroelectrics},\
  }\href {https://doi.org/10.1126/science.adp3575} {\bibfield  {journal}
  {\bibinfo  {journal} {Science}\ }\textbf {\bibinfo {volume} {385}},\ \bibinfo
  {pages} {53} (\bibinfo {year} {2024})}\BibitemShut {NoStop}%
\bibitem [{\citenamefont {Bai}\ \emph {et~al.}(2025{\natexlab{a}})\citenamefont
  {Bai}, \citenamefont {Yu}, \citenamefont {Guan}, \citenamefont {Tian},
  \citenamefont {Wang}, \citenamefont {Yao}, \citenamefont {Yang},
  \citenamefont {Lei}, \citenamefont {Xu}, \citenamefont {Liu}, \citenamefont
  {Zhu}, \citenamefont {Tu}, \citenamefont {Shen}, \citenamefont {Xiang},
  \citenamefont {Li}, \citenamefont {Xu},\ and\ \citenamefont
  {Wang}}]{bai2025subnanoseconda}%
  \BibitemOpen
  \bibfield  {author} {\bibinfo {author} {\bibfnamefont {Y.}~\bibnamefont
  {Bai}}, \bibinfo {author} {\bibfnamefont {Z.}~\bibnamefont {Yu}}, \bibinfo
  {author} {\bibfnamefont {Z.}~\bibnamefont {Guan}}, \bibinfo {author}
  {\bibfnamefont {J.}~\bibnamefont {Tian}}, \bibinfo {author} {\bibfnamefont
  {C.}~\bibnamefont {Wang}}, \bibinfo {author} {\bibfnamefont {X.}~\bibnamefont
  {Yao}}, \bibinfo {author} {\bibfnamefont {Y.}~\bibnamefont {Yang}}, \bibinfo
  {author} {\bibfnamefont {Y.}~\bibnamefont {Lei}}, \bibinfo {author}
  {\bibfnamefont {J.}~\bibnamefont {Xu}}, \bibinfo {author} {\bibfnamefont
  {C.}~\bibnamefont {Liu}}, \bibinfo {author} {\bibfnamefont {J.}~\bibnamefont
  {Zhu}}, \bibinfo {author} {\bibfnamefont {Y.}~\bibnamefont {Tu}}, \bibinfo
  {author} {\bibfnamefont {S.}~\bibnamefont {Shen}}, \bibinfo {author}
  {\bibfnamefont {H.}~\bibnamefont {Xiang}}, \bibinfo {author} {\bibfnamefont
  {X.}~\bibnamefont {Li}}, \bibinfo {author} {\bibfnamefont {C.}~\bibnamefont
  {Xu}},\ and\ \bibinfo {author} {\bibfnamefont {J.}~\bibnamefont {Wang}},\
  }\bibfield  {title} {\bibinfo {title} {Sub-nanosecond polarization switching
  with anomalous kinetics in {{vdW}} ferroelectric {{WTe2}}},\ }\href
  {https://doi.org/10.1038/s41467-025-62608-x} {\bibfield  {journal} {\bibinfo
  {journal} {Nature Communications}\ }\textbf {\bibinfo {volume} {16}},\
  \bibinfo {pages} {7221} (\bibinfo {year} {2025}{\natexlab{a}})}\BibitemShut
  {NoStop}%
\bibitem [{\citenamefont {Ji}\ \emph {et~al.}(2023)\citenamefont {Ji},
  \citenamefont {Yu}, \citenamefont {Xu},\ and\ \citenamefont
  {Xiang}}]{ji2023generalc}%
  \BibitemOpen
  \bibfield  {author} {\bibinfo {author} {\bibfnamefont {J.}~\bibnamefont
  {Ji}}, \bibinfo {author} {\bibfnamefont {G.}~\bibnamefont {Yu}}, \bibinfo
  {author} {\bibfnamefont {C.}~\bibnamefont {Xu}},\ and\ \bibinfo {author}
  {\bibfnamefont {H.~J.}\ \bibnamefont {Xiang}},\ }\bibfield  {title} {\bibinfo
  {title} {General {{Theory}} for {{Bilayer Stacking Ferroelectricity}}},\
  }\href {https://doi.org/10.1103/PhysRevLett.130.146801} {\bibfield  {journal}
  {\bibinfo  {journal} {Physical Review Letters}\ }\textbf {\bibinfo {volume}
  {130}},\ \bibinfo {pages} {146801} (\bibinfo {year} {2023})}\BibitemShut
  {NoStop}%
\bibitem [{\citenamefont {Yang}\ \emph {et~al.}(2023)\citenamefont {Yang},
  \citenamefont {Ding}, \citenamefont {Gao},\ and\ \citenamefont
  {Wu}}]{yang2023atypicala}%
  \BibitemOpen
  \bibfield  {author} {\bibinfo {author} {\bibfnamefont {L.}~\bibnamefont
  {Yang}}, \bibinfo {author} {\bibfnamefont {S.}~\bibnamefont {Ding}}, \bibinfo
  {author} {\bibfnamefont {J.}~\bibnamefont {Gao}},\ and\ \bibinfo {author}
  {\bibfnamefont {M.}~\bibnamefont {Wu}},\ }\bibfield  {title} {\bibinfo
  {title} {Atypical {{Sliding}} and {{Moir}}\textbackslash 'e
  {{Ferroelectricity}} in {{Pure Multilayer Graphene}}},\ }\href
  {https://doi.org/10.1103/PhysRevLett.131.096801} {\bibfield  {journal}
  {\bibinfo  {journal} {Physical Review Letters}\ }\textbf {\bibinfo {volume}
  {131}},\ \bibinfo {pages} {096801} (\bibinfo {year} {2023})}\BibitemShut
  {NoStop}%
\bibitem [{\citenamefont {Chen}\ \emph {et~al.}(2024)\citenamefont {Chen},
  \citenamefont {Ding}, \citenamefont {Gou},\ and\ \citenamefont
  {Zeng}}]{chen2024stronga}%
  \BibitemOpen
  \bibfield  {author} {\bibinfo {author} {\bibfnamefont {X.}~\bibnamefont
  {Chen}}, \bibinfo {author} {\bibfnamefont {X.}~\bibnamefont {Ding}}, \bibinfo
  {author} {\bibfnamefont {G.}~\bibnamefont {Gou}},\ and\ \bibinfo {author}
  {\bibfnamefont {X.~C.}\ \bibnamefont {Zeng}},\ }\bibfield  {title} {\bibinfo
  {title} {Strong {{Sliding Ferroelectricity}} and {{Interlayer Sliding
  Controllable Spintronic Effect}} in {{Two-Dimensional HgI2 Layers}}},\ }\href
  {https://doi.org/10.1021/acs.nanolett.3c04869} {\bibfield  {journal}
  {\bibinfo  {journal} {Nano Letters}\ }\textbf {\bibinfo {volume} {24}},\
  \bibinfo {pages} {3089} (\bibinfo {year} {2024})}\BibitemShut {NoStop}%
\bibitem [{\citenamefont {Guo}\ \emph {et~al.}(2025)\citenamefont {Guo},
  \citenamefont {Qian}, \citenamefont {Zhou}, \citenamefont {Wang},
  \citenamefont {Cheng},\ and\ \citenamefont {Wang}}]{guo2025slidingb}%
  \BibitemOpen
  \bibfield  {author} {\bibinfo {author} {\bibfnamefont {Z.}~\bibnamefont
  {Guo}}, \bibinfo {author} {\bibfnamefont {S.}~\bibnamefont {Qian}}, \bibinfo
  {author} {\bibfnamefont {X.}~\bibnamefont {Zhou}}, \bibinfo {author}
  {\bibfnamefont {W.}~\bibnamefont {Wang}}, \bibinfo {author} {\bibfnamefont
  {Z.}~\bibnamefont {Cheng}},\ and\ \bibinfo {author} {\bibfnamefont
  {X.}~\bibnamefont {Wang}},\ }\bibfield  {title} {\bibinfo {title} {Sliding
  ferroelectric metal with ferrimagnetism},\ }\href
  {https://doi.org/10.1038/s41467-025-67240-3} {\bibfield  {journal} {\bibinfo
  {journal} {Nature Communications}\ }\textbf {\bibinfo {volume} {17}},\
  \bibinfo {pages} {549} (\bibinfo {year} {2025})}\BibitemShut {NoStop}%
\bibitem [{\citenamefont {Jiang}\ \emph {et~al.}(2025)\citenamefont {Jiang},
  \citenamefont {Ning}, \citenamefont {Liu}, \citenamefont {Song},
  \citenamefont {Ali}, \citenamefont {Deng}, \citenamefont {Li}, \citenamefont
  {Huang}, \citenamefont {Qiu}, \citenamefont {Zhu}, \citenamefont {Fan},
  \citenamefont {Li}, \citenamefont {Qin}, \citenamefont {Xue}, \citenamefont
  {Yang}, \citenamefont {Li}, \citenamefont {Liu}, \citenamefont {Hu},
  \citenamefont {Li},\ and\ \citenamefont {Zhang}}]{jiang20252d}%
  \BibitemOpen
  \bibfield  {author} {\bibinfo {author} {\bibfnamefont {Y.}~\bibnamefont
  {Jiang}}, \bibinfo {author} {\bibfnamefont {X.}~\bibnamefont {Ning}},
  \bibinfo {author} {\bibfnamefont {R.}~\bibnamefont {Liu}}, \bibinfo {author}
  {\bibfnamefont {K.}~\bibnamefont {Song}}, \bibinfo {author} {\bibfnamefont
  {S.}~\bibnamefont {Ali}}, \bibinfo {author} {\bibfnamefont {H.}~\bibnamefont
  {Deng}}, \bibinfo {author} {\bibfnamefont {Y.}~\bibnamefont {Li}}, \bibinfo
  {author} {\bibfnamefont {B.}~\bibnamefont {Huang}}, \bibinfo {author}
  {\bibfnamefont {J.}~\bibnamefont {Qiu}}, \bibinfo {author} {\bibfnamefont
  {X.}~\bibnamefont {Zhu}}, \bibinfo {author} {\bibfnamefont {Z.}~\bibnamefont
  {Fan}}, \bibinfo {author} {\bibfnamefont {Q.}~\bibnamefont {Li}}, \bibinfo
  {author} {\bibfnamefont {C.}~\bibnamefont {Qin}}, \bibinfo {author}
  {\bibfnamefont {F.}~\bibnamefont {Xue}}, \bibinfo {author} {\bibfnamefont
  {T.}~\bibnamefont {Yang}}, \bibinfo {author} {\bibfnamefont {B.}~\bibnamefont
  {Li}}, \bibinfo {author} {\bibfnamefont {G.}~\bibnamefont {Liu}}, \bibinfo
  {author} {\bibfnamefont {W.}~\bibnamefont {Hu}}, \bibinfo {author}
  {\bibfnamefont {L.-J.}\ \bibnamefont {Li}},\ and\ \bibinfo {author}
  {\bibfnamefont {Z.}~\bibnamefont {Zhang}},\ }\bibfield  {title} {\bibinfo
  {title} {{{2D}} ferroelectric narrow-bandgap semiconductor {{Wurtzite}}' type
  {$\alpha$}-{{In2Se3}} and its silicon-compatible growth},\ }\href
  {https://doi.org/10.1038/s41467-025-62822-7} {\bibfield  {journal} {\bibinfo
  {journal} {Nature Communications}\ }\textbf {\bibinfo {volume} {16}},\
  \bibinfo {pages} {7364} (\bibinfo {year} {2025})}\BibitemShut {NoStop}%
\bibitem [{\citenamefont {Pang}\ and\ \citenamefont
  {He}(2025)}]{pang2025generalized}%
  \BibitemOpen
  \bibfield  {author} {\bibinfo {author} {\bibfnamefont {H.}~\bibnamefont
  {Pang}}\ and\ \bibinfo {author} {\bibfnamefont {L.}~\bibnamefont {He}},\
  }\bibfield  {title} {\bibinfo {title} {Generalized {{Neumann}}'s
  {{Principle}} as a {{Unified Framework}} for {{Fractional Quantum}} and
  {{Conventional Ferroelectricity}}},\ }\href
  {https://doi.org/10.1103/trhd-kxm1} {\bibfield  {journal} {\bibinfo
  {journal} {Physical Review Letters}\ }\textbf {\bibinfo {volume} {135}},\
  \bibinfo {pages} {116402} (\bibinfo {year} {2025})}\BibitemShut {NoStop}%
\bibitem [{\citenamefont {Yu}\ \emph {et~al.}(2025)\citenamefont {Yu},
  \citenamefont {Ji}, \citenamefont {Chen}, \citenamefont {Xu},\ and\
  \citenamefont {Xiang}}]{yu2025symmetry}%
  \BibitemOpen
  \bibfield  {author} {\bibinfo {author} {\bibfnamefont {G.}~\bibnamefont
  {Yu}}, \bibinfo {author} {\bibfnamefont {J.}~\bibnamefont {Ji}}, \bibinfo
  {author} {\bibfnamefont {Y.}~\bibnamefont {Chen}}, \bibinfo {author}
  {\bibfnamefont {C.}~\bibnamefont {Xu}},\ and\ \bibinfo {author}
  {\bibfnamefont {H.~J.}\ \bibnamefont {Xiang}},\ }\bibfield  {title} {\bibinfo
  {title} {Symmetry {{Strategy}} for {{Rapid Discovery}} of {{Abundant
  Fractional Quantum Ferroelectrics}}},\ }\href
  {https://doi.org/10.1103/PhysRevLett.134.016801} {\bibfield  {journal}
  {\bibinfo  {journal} {Physical Review Letters}\ }\textbf {\bibinfo {volume}
  {134}},\ \bibinfo {pages} {016801} (\bibinfo {year} {2025})}\BibitemShut
  {NoStop}%
\bibitem [{\citenamefont {Dong}\ \emph {et~al.}(2026)\citenamefont {Dong},
  \citenamefont {Liu}, \citenamefont {Dai}, \citenamefont {Guo}, \citenamefont
  {Xiang},\ and\ \citenamefont {Gong}}]{dong2026fractionala}%
  \BibitemOpen
  \bibfield  {author} {\bibinfo {author} {\bibfnamefont {M.~Q.}\ \bibnamefont
  {Dong}}, \bibinfo {author} {\bibfnamefont {B.}~\bibnamefont {Liu}}, \bibinfo
  {author} {\bibfnamefont {Z.~H.}\ \bibnamefont {Dai}}, \bibinfo {author}
  {\bibfnamefont {Z.-X.}\ \bibnamefont {Guo}}, \bibinfo {author} {\bibfnamefont
  {H.}~\bibnamefont {Xiang}},\ and\ \bibinfo {author} {\bibfnamefont {X.-G.}\
  \bibnamefont {Gong}},\ }\bibfield  {title} {\bibinfo {title} {Fractional
  {{Quantum Multiferroics}} from {{Coupling}} of {{Fractional Quantum
  Ferroelectricity}} and {{Altermagnetism}}},\ }\href
  {https://doi.org/10.1103/twhq-32db} {\bibfield  {journal} {\bibinfo
  {journal} {Physical Review Letters}\ }\textbf {\bibinfo {volume} {136}},\
  \bibinfo {pages} {136702} (\bibinfo {year} {2026})}\BibitemShut {NoStop}%
\bibitem [{\citenamefont {Vanderbilt}(2018)}]{vanderbilt2018berry}%
  \BibitemOpen
  \bibfield  {author} {\bibinfo {author} {\bibfnamefont {D.}~\bibnamefont
  {Vanderbilt}},\ }\href@noop {} {\emph {\bibinfo {title} {Berry {{Phases}} in
  {{Electronic Structure Theory}}: {{Electric Polarization}}, {{Orbital
  Magnetization}} and {{Topological Insulators}}}}}\ (\bibinfo  {publisher}
  {Cambridge University Press},\ \bibinfo {year} {2018})\BibitemShut {NoStop}%
\bibitem [{\citenamefont {Fox}\ \emph {et~al.}(2025)\citenamefont {Fox},
  \citenamefont {Mella}, \citenamefont {Rollins}, \citenamefont {He},
  \citenamefont {Mao}, \citenamefont {Jiang}, \citenamefont {Drew},
  \citenamefont {Ma}, \citenamefont {Taniguchi}, \citenamefont {Watanabe},
  \citenamefont {Wang}, \citenamefont {Rhodes}, \citenamefont
  {{Barraza-Lopez}},\ and\ \citenamefont {Xiao}}]{fox2025sliding}%
  \BibitemOpen
  \bibfield  {author} {\bibinfo {author} {\bibfnamefont {C.}~\bibnamefont
  {Fox}}, \bibinfo {author} {\bibfnamefont {J.~D.}\ \bibnamefont {Mella}},
  \bibinfo {author} {\bibfnamefont {J.}~\bibnamefont {Rollins}}, \bibinfo
  {author} {\bibfnamefont {Y.}~\bibnamefont {He}}, \bibinfo {author}
  {\bibfnamefont {Y.}~\bibnamefont {Mao}}, \bibinfo {author} {\bibfnamefont
  {H.}~\bibnamefont {Jiang}}, \bibinfo {author} {\bibfnamefont
  {A.}~\bibnamefont {Drew}}, \bibinfo {author} {\bibfnamefont {H.}~\bibnamefont
  {Ma}}, \bibinfo {author} {\bibfnamefont {T.}~\bibnamefont {Taniguchi}},
  \bibinfo {author} {\bibfnamefont {K.}~\bibnamefont {Watanabe}}, \bibinfo
  {author} {\bibfnamefont {Y.}~\bibnamefont {Wang}}, \bibinfo {author}
  {\bibfnamefont {D.}~\bibnamefont {Rhodes}}, \bibinfo {author} {\bibfnamefont
  {S.}~\bibnamefont {{Barraza-Lopez}}},\ and\ \bibinfo {author} {\bibfnamefont
  {J.}~\bibnamefont {Xiao}},\ }\href
  {https://doi.org/10.48550/arXiv.2510.03220} {\bibinfo {title} {Sliding
  multiferroicity in hexagonal stacked {{CrI3}}}} (\bibinfo {year} {2025}),\
  \Eprint {https://arxiv.org/abs/2510.03220} {arXiv:2510.03220
  [cond-mat.mtrl-sci]} \BibitemShut {NoStop}%
\bibitem [{\citenamefont {Ma}\ \emph {et~al.}(2024)\citenamefont {Ma},
  \citenamefont {Luo},\ and\ \citenamefont {Zheng}}]{ma2024strain}%
  \BibitemOpen
  \bibfield  {author} {\bibinfo {author} {\bibfnamefont {J.}~\bibnamefont
  {Ma}}, \bibinfo {author} {\bibfnamefont {X.}~\bibnamefont {Luo}},\ and\
  \bibinfo {author} {\bibfnamefont {Y.}~\bibnamefont {Zheng}},\ }\bibfield
  {title} {\bibinfo {title} {Strain engineering the spin-valley coupling of the
  {{R-stacking}} sliding ferroelectric bilayer {{2H-VX2}} ({{X}} = {{S}},
  {{Se}}, {{Te}})},\ }\href {https://doi.org/10.1038/s41524-024-01288-5}
  {\bibfield  {journal} {\bibinfo  {journal} {npj Computational Materials}\
  }\textbf {\bibinfo {volume} {10}},\ \bibinfo {pages} {102} (\bibinfo {year}
  {2024})}\BibitemShut {NoStop}%
\bibitem [{\citenamefont {Sheng}\ \emph {et~al.}(2025)\citenamefont {Sheng},
  \citenamefont {Zhang}, \citenamefont {Liu},\ and\ \citenamefont
  {Wu}}]{sheng2025ubiquitous}%
  \BibitemOpen
  \bibfield  {author} {\bibinfo {author} {\bibfnamefont {Y.}~\bibnamefont
  {Sheng}}, \bibinfo {author} {\bibfnamefont {J.}~\bibnamefont {Zhang}},
  \bibinfo {author} {\bibfnamefont {J.}~\bibnamefont {Liu}},\ and\ \bibinfo
  {author} {\bibfnamefont {M.}~\bibnamefont {Wu}},\ }\bibfield  {title}
  {\bibinfo {title} {Ubiquitous van der {{Waals}} altermagnetism with
  sliding/moire ferroelectricity},\ }\href
  {https://doi.org/10.1007/s11433-025-2698-5} {\bibfield  {journal} {\bibinfo
  {journal} {Science China Physics, Mechanics \& Astronomy}\ }\textbf {\bibinfo
  {volume} {68}},\ \bibinfo {pages} {297511} (\bibinfo {year}
  {2025})}\BibitemShut {NoStop}%
\bibitem [{\citenamefont {Zhang}\ \emph {et~al.}(2017)\citenamefont {Zhang},
  \citenamefont {Niu}, \citenamefont {Yang}, \citenamefont {Gong},
  \citenamefont {Ji}, \citenamefont {Shi}, \citenamefont {Fang}, \citenamefont
  {Jiang}, \citenamefont {Li}, \citenamefont {Zhou}, \citenamefont {Gu},
  \citenamefont {Wu},\ and\ \citenamefont {Zhang}}]{zhang2017van}%
  \BibitemOpen
  \bibfield  {author} {\bibinfo {author} {\bibfnamefont {Z.}~\bibnamefont
  {Zhang}}, \bibinfo {author} {\bibfnamefont {J.}~\bibnamefont {Niu}}, \bibinfo
  {author} {\bibfnamefont {P.}~\bibnamefont {Yang}}, \bibinfo {author}
  {\bibfnamefont {Y.}~\bibnamefont {Gong}}, \bibinfo {author} {\bibfnamefont
  {Q.}~\bibnamefont {Ji}}, \bibinfo {author} {\bibfnamefont {J.}~\bibnamefont
  {Shi}}, \bibinfo {author} {\bibfnamefont {Q.}~\bibnamefont {Fang}}, \bibinfo
  {author} {\bibfnamefont {S.}~\bibnamefont {Jiang}}, \bibinfo {author}
  {\bibfnamefont {H.}~\bibnamefont {Li}}, \bibinfo {author} {\bibfnamefont
  {X.}~\bibnamefont {Zhou}}, \bibinfo {author} {\bibfnamefont {L.}~\bibnamefont
  {Gu}}, \bibinfo {author} {\bibfnamefont {X.}~\bibnamefont {Wu}},\ and\
  \bibinfo {author} {\bibfnamefont {Y.}~\bibnamefont {Zhang}},\ }\bibfield
  {title} {\bibinfo {title} {Van der {{Waals Epitaxial Growth}} of {{2D
  Metallic Vanadium Diselenide Single Crystals}} and their {{Extra-High
  Electrical Conductivity}}},\ }\href {https://doi.org/10.1002/adma.201702359}
  {\bibfield  {journal} {\bibinfo  {journal} {Advanced Materials}\ }\textbf
  {\bibinfo {volume} {29}},\ \bibinfo {pages} {1702359} (\bibinfo {year}
  {2017})}\BibitemShut {NoStop}%
\bibitem [{\citenamefont {Bonilla}\ \emph {et~al.}(2018)\citenamefont
  {Bonilla}, \citenamefont {Kolekar}, \citenamefont {Ma}, \citenamefont {Diaz},
  \citenamefont {Kalappattil}, \citenamefont {Das}, \citenamefont {Eggers},
  \citenamefont {Gutierrez}, \citenamefont {Phan},\ and\ \citenamefont
  {Batzill}}]{bonilla2018strong}%
  \BibitemOpen
  \bibfield  {author} {\bibinfo {author} {\bibfnamefont {M.}~\bibnamefont
  {Bonilla}}, \bibinfo {author} {\bibfnamefont {S.}~\bibnamefont {Kolekar}},
  \bibinfo {author} {\bibfnamefont {Y.}~\bibnamefont {Ma}}, \bibinfo {author}
  {\bibfnamefont {H.~C.}\ \bibnamefont {Diaz}}, \bibinfo {author}
  {\bibfnamefont {V.}~\bibnamefont {Kalappattil}}, \bibinfo {author}
  {\bibfnamefont {R.}~\bibnamefont {Das}}, \bibinfo {author} {\bibfnamefont
  {T.}~\bibnamefont {Eggers}}, \bibinfo {author} {\bibfnamefont {H.~R.}\
  \bibnamefont {Gutierrez}}, \bibinfo {author} {\bibfnamefont {M.-H.}\
  \bibnamefont {Phan}},\ and\ \bibinfo {author} {\bibfnamefont
  {M.}~\bibnamefont {Batzill}},\ }\bibfield  {title} {\bibinfo {title} {Strong
  room-temperature ferromagnetism in {{VSe2}} monolayers on van der {{Waals}}
  substrates},\ }\href {https://doi.org/10.1038/s41565-018-0063-9} {\bibfield
  {journal} {\bibinfo  {journal} {Nature Nanotechnology}\ }\textbf {\bibinfo
  {volume} {13}},\ \bibinfo {pages} {289} (\bibinfo {year} {2018})}\BibitemShut
  {NoStop}%
\bibitem [{\citenamefont {Liu}\ \emph {et~al.}(2019{\natexlab{a}})\citenamefont
  {Liu}, \citenamefont {Xue}, \citenamefont {Shi}, \citenamefont {Guzman},
  \citenamefont {Zhang}, \citenamefont {Zhou}, \citenamefont {He},
  \citenamefont {Bian}, \citenamefont {Wu}, \citenamefont {Ma}, \citenamefont
  {Chen}, \citenamefont {Yan}, \citenamefont {Yang}, \citenamefont {Shen},
  \citenamefont {Zhou}, \citenamefont {Bao},\ and\ \citenamefont
  {Gao}}]{liu2019observation}%
  \BibitemOpen
  \bibfield  {author} {\bibinfo {author} {\bibfnamefont {H.}~\bibnamefont
  {Liu}}, \bibinfo {author} {\bibfnamefont {Y.}~\bibnamefont {Xue}}, \bibinfo
  {author} {\bibfnamefont {J.-A.}\ \bibnamefont {Shi}}, \bibinfo {author}
  {\bibfnamefont {R.~A.}\ \bibnamefont {Guzman}}, \bibinfo {author}
  {\bibfnamefont {P.}~\bibnamefont {Zhang}}, \bibinfo {author} {\bibfnamefont
  {Z.}~\bibnamefont {Zhou}}, \bibinfo {author} {\bibfnamefont {Y.}~\bibnamefont
  {He}}, \bibinfo {author} {\bibfnamefont {C.}~\bibnamefont {Bian}}, \bibinfo
  {author} {\bibfnamefont {L.}~\bibnamefont {Wu}}, \bibinfo {author}
  {\bibfnamefont {R.}~\bibnamefont {Ma}}, \bibinfo {author} {\bibfnamefont
  {J.}~\bibnamefont {Chen}}, \bibinfo {author} {\bibfnamefont {J.}~\bibnamefont
  {Yan}}, \bibinfo {author} {\bibfnamefont {H.}~\bibnamefont {Yang}}, \bibinfo
  {author} {\bibfnamefont {C.-M.}\ \bibnamefont {Shen}}, \bibinfo {author}
  {\bibfnamefont {W.}~\bibnamefont {Zhou}}, \bibinfo {author} {\bibfnamefont
  {L.}~\bibnamefont {Bao}},\ and\ \bibinfo {author} {\bibfnamefont {H.-J.}\
  \bibnamefont {Gao}},\ }\bibfield  {title} {\bibinfo {title} {Observation of
  the {{Kondo Effect}} in {{Multilayer Single-Crystalline VTe2 Nanoplates}}},\
  }\href {https://doi.org/10.1021/acs.nanolett.9b03100} {\bibfield  {journal}
  {\bibinfo  {journal} {Nano Letters}\ }\textbf {\bibinfo {volume} {19}},\
  \bibinfo {pages} {8572} (\bibinfo {year} {2019}{\natexlab{a}})}\BibitemShut
  {NoStop}%
\bibitem [{\citenamefont {Liu}\ \emph {et~al.}(2019{\natexlab{b}})\citenamefont
  {Liu}, \citenamefont {Bao}, \citenamefont {Zhou}, \citenamefont {Che},
  \citenamefont {Zhang}, \citenamefont {Bian}, \citenamefont {Ma},
  \citenamefont {Wu}, \citenamefont {Yang}, \citenamefont {Li}, \citenamefont
  {Gu}, \citenamefont {Shen}, \citenamefont {Du},\ and\ \citenamefont
  {Gao}}]{liu2019quasi2da}%
  \BibitemOpen
  \bibfield  {author} {\bibinfo {author} {\bibfnamefont {H.}~\bibnamefont
  {Liu}}, \bibinfo {author} {\bibfnamefont {L.}~\bibnamefont {Bao}}, \bibinfo
  {author} {\bibfnamefont {Z.}~\bibnamefont {Zhou}}, \bibinfo {author}
  {\bibfnamefont {B.}~\bibnamefont {Che}}, \bibinfo {author} {\bibfnamefont
  {R.}~\bibnamefont {Zhang}}, \bibinfo {author} {\bibfnamefont
  {C.}~\bibnamefont {Bian}}, \bibinfo {author} {\bibfnamefont {R.}~\bibnamefont
  {Ma}}, \bibinfo {author} {\bibfnamefont {L.}~\bibnamefont {Wu}}, \bibinfo
  {author} {\bibfnamefont {H.}~\bibnamefont {Yang}}, \bibinfo {author}
  {\bibfnamefont {J.}~\bibnamefont {Li}}, \bibinfo {author} {\bibfnamefont
  {C.}~\bibnamefont {Gu}}, \bibinfo {author} {\bibfnamefont {C.-M.}\
  \bibnamefont {Shen}}, \bibinfo {author} {\bibfnamefont {S.}~\bibnamefont
  {Du}},\ and\ \bibinfo {author} {\bibfnamefont {H.-J.}\ \bibnamefont {Gao}},\
  }\bibfield  {title} {\bibinfo {title} {Quasi-{{2D Transport}} and {{Weak
  Antilocalization Effect}} in {{Few-layered VSe2}}},\ }\href
  {https://doi.org/10.1021/acs.nanolett.9b01412} {\bibfield  {journal}
  {\bibinfo  {journal} {Nano Letters}\ }\textbf {\bibinfo {volume} {19}},\
  \bibinfo {pages} {4551} (\bibinfo {year} {2019}{\natexlab{b}})}\BibitemShut
  {NoStop}%
\bibitem [{\citenamefont {Li}\ \emph {et~al.}(2020)\citenamefont {Li},
  \citenamefont {Wang}, \citenamefont {Kan}, \citenamefont {He}, \citenamefont
  {Li}, \citenamefont {Hao}, \citenamefont {Zhao}, \citenamefont {Wu},
  \citenamefont {Jin},\ and\ \citenamefont {Cui}}]{li2020structural}%
  \BibitemOpen
  \bibfield  {author} {\bibinfo {author} {\bibfnamefont {D.}~\bibnamefont
  {Li}}, \bibinfo {author} {\bibfnamefont {X.}~\bibnamefont {Wang}}, \bibinfo
  {author} {\bibfnamefont {C.-m.}\ \bibnamefont {Kan}}, \bibinfo {author}
  {\bibfnamefont {D.}~\bibnamefont {He}}, \bibinfo {author} {\bibfnamefont
  {Z.}~\bibnamefont {Li}}, \bibinfo {author} {\bibfnamefont {Q.}~\bibnamefont
  {Hao}}, \bibinfo {author} {\bibfnamefont {H.}~\bibnamefont {Zhao}}, \bibinfo
  {author} {\bibfnamefont {C.}~\bibnamefont {Wu}}, \bibinfo {author}
  {\bibfnamefont {C.}~\bibnamefont {Jin}},\ and\ \bibinfo {author}
  {\bibfnamefont {X.}~\bibnamefont {Cui}},\ }\bibfield  {title} {\bibinfo
  {title} {Structural {{Phase Transition}} of {{Multilayer VSe2}}},\ }\href
  {https://doi.org/10.1021/acsami.0c04449} {\bibfield  {journal} {\bibinfo
  {journal} {ACS Applied Materials \& Interfaces}\ }\textbf {\bibinfo {volume}
  {12}},\ \bibinfo {pages} {25143} (\bibinfo {year} {2020})}\BibitemShut
  {NoStop}%
\bibitem [{\citenamefont {Su}\ \emph {et~al.}(2020)\citenamefont {Su},
  \citenamefont {Wang}, \citenamefont {Li}, \citenamefont {Wang}, \citenamefont
  {Chen}, \citenamefont {Luo}, \citenamefont {Han}, \citenamefont {Wang},
  \citenamefont {Li},\ and\ \citenamefont {Zhai}}]{su2020submillimeterscale}%
  \BibitemOpen
  \bibfield  {author} {\bibinfo {author} {\bibfnamefont {J.}~\bibnamefont
  {Su}}, \bibinfo {author} {\bibfnamefont {M.}~\bibnamefont {Wang}}, \bibinfo
  {author} {\bibfnamefont {Y.}~\bibnamefont {Li}}, \bibinfo {author}
  {\bibfnamefont {F.}~\bibnamefont {Wang}}, \bibinfo {author} {\bibfnamefont
  {Q.}~\bibnamefont {Chen}}, \bibinfo {author} {\bibfnamefont {P.}~\bibnamefont
  {Luo}}, \bibinfo {author} {\bibfnamefont {J.}~\bibnamefont {Han}}, \bibinfo
  {author} {\bibfnamefont {S.}~\bibnamefont {Wang}}, \bibinfo {author}
  {\bibfnamefont {H.}~\bibnamefont {Li}},\ and\ \bibinfo {author}
  {\bibfnamefont {T.}~\bibnamefont {Zhai}},\ }\bibfield  {title} {\bibinfo
  {title} {Sub-{{Millimeter-Scale Monolayer}} p-{{Type H-Phase VS2}}},\ }\href
  {https://doi.org/10.1002/adfm.202000240} {\bibfield  {journal} {\bibinfo
  {journal} {Advanced Functional Materials}\ }\textbf {\bibinfo {volume}
  {30}},\ \bibinfo {pages} {2000240} (\bibinfo {year} {2020})}\BibitemShut
  {NoStop}%
\bibitem [{\citenamefont {Zhu}\ \emph {et~al.}(2022)\citenamefont {Zhu},
  \citenamefont {Liu}, \citenamefont {Wu}, \citenamefont {Li}, \citenamefont
  {Shi}, \citenamefont {Liu}, \citenamefont {Qian}, \citenamefont {Zheng},
  \citenamefont {Huang}, \citenamefont {Lin}, \citenamefont {Wang},
  \citenamefont {Chen}, \citenamefont {Zhou}, \citenamefont {Sun},
  \citenamefont {Wang},\ and\ \citenamefont {Gao}}]{zhu2022charge}%
  \BibitemOpen
  \bibfield  {author} {\bibinfo {author} {\bibfnamefont {Z.-L.}\ \bibnamefont
  {Zhu}}, \bibinfo {author} {\bibfnamefont {Z.-L.}\ \bibnamefont {Liu}},
  \bibinfo {author} {\bibfnamefont {X.}~\bibnamefont {Wu}}, \bibinfo {author}
  {\bibfnamefont {X.-Y.}\ \bibnamefont {Li}}, \bibinfo {author} {\bibfnamefont
  {J.-A.}\ \bibnamefont {Shi}}, \bibinfo {author} {\bibfnamefont
  {C.}~\bibnamefont {Liu}}, \bibinfo {author} {\bibfnamefont {G.-J.}\
  \bibnamefont {Qian}}, \bibinfo {author} {\bibfnamefont {Q.}~\bibnamefont
  {Zheng}}, \bibinfo {author} {\bibfnamefont {L.}~\bibnamefont {Huang}},
  \bibinfo {author} {\bibfnamefont {X.}~\bibnamefont {Lin}}, \bibinfo {author}
  {\bibfnamefont {J.-O.}\ \bibnamefont {Wang}}, \bibinfo {author}
  {\bibfnamefont {H.}~\bibnamefont {Chen}}, \bibinfo {author} {\bibfnamefont
  {W.}~\bibnamefont {Zhou}}, \bibinfo {author} {\bibfnamefont {J.-T.}\
  \bibnamefont {Sun}}, \bibinfo {author} {\bibfnamefont {Y.-L.}\ \bibnamefont
  {Wang}},\ and\ \bibinfo {author} {\bibfnamefont {H.-J.}\ \bibnamefont
  {Gao}},\ }\bibfield  {title} {\bibinfo {title} {Charge density wave states in
  phase-engineered monolayer {{VTe2}}},\ }\href
  {https://doi.org/10.1088/1674-1056/ac6739} {\bibfield  {journal} {\bibinfo
  {journal} {Chinese Physics B}\ }\textbf {\bibinfo {volume} {31}},\ \bibinfo
  {pages} {077101} (\bibinfo {year} {2022})}\BibitemShut {NoStop}%
\bibitem [{Han()}]{Hansupplemental}%
  \BibitemOpen
  \href@noop {} {\ }\bibinfo {note} {See Supplemental Material for calculation
  methods and analysis of sliding spin current in other magnetic sliding
  ferroelectrics
  \cite{kresse1996efficient,kresse1996efficiency,blochl1994projector,kresse1999ultrasoft,perdew1996generalized,perdew1998perdew,monkhorst1976speciale,grimme2010consistent,anisimov1991band,resta1992theory,king-smith1993theory,resta1994macroscopic,vanderbilt2018berry,ma2024strain,sheng2025ubiquitous,feng2024van,zhang2023layerpolarized,li2024sliding,zhong2023theoreticala,pan2025stackingtunable}.}\BibitemShut
  {Stop}%
\bibitem [{\citenamefont {{van Leuken}}\ and\ \citenamefont {{de
  Groot}}(1995)}]{vanleuken1995halfmetallic}%
  \BibitemOpen
  \bibfield  {author} {\bibinfo {author} {\bibfnamefont {H.}~\bibnamefont {{van
  Leuken}}}\ and\ \bibinfo {author} {\bibfnamefont {R.~A.}\ \bibnamefont {{de
  Groot}}},\ }\bibfield  {title} {\bibinfo {title} {Half-{{Metallic
  Antiferromagnets}}},\ }\href {https://doi.org/10.1103/PhysRevLett.74.1171}
  {\bibfield  {journal} {\bibinfo  {journal} {Physical Review Letters}\
  }\textbf {\bibinfo {volume} {74}},\ \bibinfo {pages} {1171} (\bibinfo {year}
  {1995})}\BibitemShut {NoStop}%
\bibitem [{\citenamefont {Kawamura}\ \emph {et~al.}(2024)\citenamefont
  {Kawamura}, \citenamefont {Yoshimi}, \citenamefont {Hashimoto}, \citenamefont
  {Kobayashi},\ and\ \citenamefont {Misawa}}]{kawamura2024compensateda}%
  \BibitemOpen
  \bibfield  {author} {\bibinfo {author} {\bibfnamefont {T.}~\bibnamefont
  {Kawamura}}, \bibinfo {author} {\bibfnamefont {K.}~\bibnamefont {Yoshimi}},
  \bibinfo {author} {\bibfnamefont {K.}~\bibnamefont {Hashimoto}}, \bibinfo
  {author} {\bibfnamefont {A.}~\bibnamefont {Kobayashi}},\ and\ \bibinfo
  {author} {\bibfnamefont {T.}~\bibnamefont {Misawa}},\ }\bibfield  {title}
  {\bibinfo {title} {Compensated {{Ferrimagnets}} with {{Colossal Spin
  Splitting}} in {{Organic Compounds}}},\ }\href
  {https://doi.org/10.1103/PhysRevLett.132.156502} {\bibfield  {journal}
  {\bibinfo  {journal} {Physical Review Letters}\ }\textbf {\bibinfo {volume}
  {132}},\ \bibinfo {pages} {156502} (\bibinfo {year} {2024})}\BibitemShut
  {NoStop}%
\bibitem [{\citenamefont {Yuan}\ \emph {et~al.}(2024)\citenamefont {Yuan},
  \citenamefont {Georgescu},\ and\ \citenamefont
  {Rondinelli}}]{yuan2024nonrelativistic}%
  \BibitemOpen
  \bibfield  {author} {\bibinfo {author} {\bibfnamefont {L.-D.}\ \bibnamefont
  {Yuan}}, \bibinfo {author} {\bibfnamefont {A.~B.}\ \bibnamefont
  {Georgescu}},\ and\ \bibinfo {author} {\bibfnamefont {J.~M.}\ \bibnamefont
  {Rondinelli}},\ }\bibfield  {title} {\bibinfo {title} {Nonrelativistic {{Spin
  Splitting}} at the {{Brillouin Zone Center}} in {{Compensated Magnets}}},\
  }\href {https://doi.org/10.1103/PhysRevLett.133.216701} {\bibfield  {journal}
  {\bibinfo  {journal} {Physical Review Letters}\ }\textbf {\bibinfo {volume}
  {133}},\ \bibinfo {pages} {216701} (\bibinfo {year} {2024})}\BibitemShut
  {NoStop}%
\bibitem [{\citenamefont {Liu}\ \emph {et~al.}(2025)\citenamefont {Liu},
  \citenamefont {Guo}, \citenamefont {Li},\ and\ \citenamefont
  {Liu}}]{liu2025twodimensionalb}%
  \BibitemOpen
  \bibfield  {author} {\bibinfo {author} {\bibfnamefont {Y.}~\bibnamefont
  {Liu}}, \bibinfo {author} {\bibfnamefont {S.-D.}\ \bibnamefont {Guo}},
  \bibinfo {author} {\bibfnamefont {Y.}~\bibnamefont {Li}},\ and\ \bibinfo
  {author} {\bibfnamefont {C.-C.}\ \bibnamefont {Liu}},\ }\bibfield  {title}
  {\bibinfo {title} {Two-{{Dimensional Fully Compensated Ferrimagnetism}}},\
  }\href {https://doi.org/10.1103/PhysRevLett.134.116703} {\bibfield  {journal}
  {\bibinfo  {journal} {Physical Review Letters}\ }\textbf {\bibinfo {volume}
  {134}},\ \bibinfo {pages} {116703} (\bibinfo {year} {2025})}\BibitemShut
  {NoStop}%
\bibitem [{\citenamefont {Bai}\ \emph {et~al.}(2025{\natexlab{b}})\citenamefont
  {Bai}, \citenamefont {Zhang}, \citenamefont {Feng},\ and\ \citenamefont
  {Yao}}]{bai2025anomalous}%
  \BibitemOpen
  \bibfield  {author} {\bibinfo {author} {\bibfnamefont {L.}~\bibnamefont
  {Bai}}, \bibinfo {author} {\bibfnamefont {R.-W.}\ \bibnamefont {Zhang}},
  \bibinfo {author} {\bibfnamefont {W.}~\bibnamefont {Feng}},\ and\ \bibinfo
  {author} {\bibfnamefont {Y.}~\bibnamefont {Yao}},\ }\bibfield  {title}
  {\bibinfo {title} {Anomalous {{Hall Effect}} in {{Type IV 2D Collinear
  Magnets}}},\ }\href {https://doi.org/10.1103/rn1l-d6cq} {\bibfield  {journal}
  {\bibinfo  {journal} {Physical Review Letters}\ }\textbf {\bibinfo {volume}
  {135}},\ \bibinfo {pages} {036702} (\bibinfo {year}
  {2025}{\natexlab{b}})}\BibitemShut {NoStop}%
\bibitem [{\citenamefont {Tian}\ \emph {et~al.}(2026)\citenamefont {Tian},
  \citenamefont {Cui}, \citenamefont {Zhang}, \citenamefont {Duan},
  \citenamefont {Feng},\ and\ \citenamefont {Zhang}}]{tian2026symmetry}%
  \BibitemOpen
  \bibfield  {author} {\bibinfo {author} {\bibfnamefont {M.}~\bibnamefont
  {Tian}}, \bibinfo {author} {\bibfnamefont {C.}~\bibnamefont {Cui}}, \bibinfo
  {author} {\bibfnamefont {Z.}~\bibnamefont {Zhang}}, \bibinfo {author}
  {\bibfnamefont {J.}~\bibnamefont {Duan}}, \bibinfo {author} {\bibfnamefont
  {W.}~\bibnamefont {Feng}},\ and\ \bibinfo {author} {\bibfnamefont {R.-W.}\
  \bibnamefont {Zhang}},\ }\bibfield  {title} {\bibinfo {title} {Symmetry
  {{Classification}} of {{Altermagnetism}} and {{Emergence}} of {{Type-IV
  Magnetism}} in {{Two Dimensions}}},\ }\href
  {https://doi.org/10.1103/jp95-17sz} {\bibfield  {journal} {\bibinfo
  {journal} {Physical Review Letters}\ }\textbf {\bibinfo {volume} {136}},\
  \bibinfo {pages} {206701} (\bibinfo {year} {2026})}\BibitemShut {NoStop}%
\bibitem [{\citenamefont {Yang}\ and\ \citenamefont
  {Meng}(2024)}]{yang2024lightinduced}%
  \BibitemOpen
  \bibfield  {author} {\bibinfo {author} {\bibfnamefont {Q.}~\bibnamefont
  {Yang}}\ and\ \bibinfo {author} {\bibfnamefont {S.}~\bibnamefont {Meng}},\
  }\bibfield  {title} {\bibinfo {title} {Light-{{Induced Complete Reversal}} of
  {{Ferroelectric Polarization}} in {{Sliding Ferroelectrics}}},\ }\href
  {https://doi.org/10.1103/PhysRevLett.133.136902} {\bibfield  {journal}
  {\bibinfo  {journal} {Physical Review Letters}\ }\textbf {\bibinfo {volume}
  {133}},\ \bibinfo {pages} {136902} (\bibinfo {year} {2024})}\BibitemShut
  {NoStop}%
\bibitem [{\citenamefont {Kresse}\ and\ \citenamefont
  {Furthm{\"u}ller}(1996{\natexlab{a}})}]{kresse1996efficient}%
  \BibitemOpen
  \bibfield  {author} {\bibinfo {author} {\bibfnamefont {G.}~\bibnamefont
  {Kresse}}\ and\ \bibinfo {author} {\bibfnamefont {J.}~\bibnamefont
  {Furthm{\"u}ller}},\ }\bibfield  {title} {\bibinfo {title} {Efficient
  iterative schemes for ab initio total-energy calculations using a plane-wave
  basis set},\ }\href {https://doi.org/10.1103/PhysRevB.54.11169} {\bibfield
  {journal} {\bibinfo  {journal} {Physical Review B}\ }\textbf {\bibinfo
  {volume} {54}},\ \bibinfo {pages} {11169} (\bibinfo {year}
  {1996}{\natexlab{a}})}\BibitemShut {NoStop}%
\bibitem [{\citenamefont {Kresse}\ and\ \citenamefont
  {Furthm{\"u}ller}(1996{\natexlab{b}})}]{kresse1996efficiency}%
  \BibitemOpen
  \bibfield  {author} {\bibinfo {author} {\bibfnamefont {G.}~\bibnamefont
  {Kresse}}\ and\ \bibinfo {author} {\bibfnamefont {J.}~\bibnamefont
  {Furthm{\"u}ller}},\ }\bibfield  {title} {\bibinfo {title} {Efficiency of
  ab-initio total energy calculations for metals and semiconductors using a
  plane-wave basis set},\ }\href {https://doi.org/10.1016/0927-0256(96)00008-0}
  {\bibfield  {journal} {\bibinfo  {journal} {Computational Materials Science}\
  }\textbf {\bibinfo {volume} {6}},\ \bibinfo {pages} {15} (\bibinfo {year}
  {1996}{\natexlab{b}})}\BibitemShut {NoStop}%
\bibitem [{\citenamefont {Bl{\"o}chl}(1994)}]{blochl1994projector}%
  \BibitemOpen
  \bibfield  {author} {\bibinfo {author} {\bibfnamefont {P.~E.}\ \bibnamefont
  {Bl{\"o}chl}},\ }\bibfield  {title} {\bibinfo {title} {Projector
  augmented-wave method},\ }\href {https://doi.org/10.1103/PhysRevB.50.17953}
  {\bibfield  {journal} {\bibinfo  {journal} {Physical Review B}\ }\textbf
  {\bibinfo {volume} {50}},\ \bibinfo {pages} {17953} (\bibinfo {year}
  {1994})}\BibitemShut {NoStop}%
\bibitem [{\citenamefont {Kresse}\ and\ \citenamefont
  {Joubert}(1999)}]{kresse1999ultrasoft}%
  \BibitemOpen
  \bibfield  {author} {\bibinfo {author} {\bibfnamefont {G.}~\bibnamefont
  {Kresse}}\ and\ \bibinfo {author} {\bibfnamefont {D.}~\bibnamefont
  {Joubert}},\ }\bibfield  {title} {\bibinfo {title} {From ultrasoft
  pseudopotentials to the projector augmented-wave method},\ }\href
  {https://doi.org/10.1103/PhysRevB.59.1758} {\bibfield  {journal} {\bibinfo
  {journal} {Physical Review B}\ }\textbf {\bibinfo {volume} {59}},\ \bibinfo
  {pages} {1758} (\bibinfo {year} {1999})}\BibitemShut {NoStop}%
\bibitem [{\citenamefont {Perdew}\ \emph {et~al.}(1996)\citenamefont {Perdew},
  \citenamefont {Burke},\ and\ \citenamefont
  {Ernzerhof}}]{perdew1996generalized}%
  \BibitemOpen
  \bibfield  {author} {\bibinfo {author} {\bibfnamefont {J.~P.}\ \bibnamefont
  {Perdew}}, \bibinfo {author} {\bibfnamefont {K.}~\bibnamefont {Burke}},\ and\
  \bibinfo {author} {\bibfnamefont {M.}~\bibnamefont {Ernzerhof}},\ }\bibfield
  {title} {\bibinfo {title} {Generalized {{Gradient Approximation Made
  Simple}}},\ }\href {https://doi.org/10.1103/PhysRevLett.77.3865} {\bibfield
  {journal} {\bibinfo  {journal} {Physical Review Letters}\ }\textbf {\bibinfo
  {volume} {77}},\ \bibinfo {pages} {3865} (\bibinfo {year}
  {1996})}\BibitemShut {NoStop}%
\bibitem [{\citenamefont {Perdew}\ \emph {et~al.}(1998)\citenamefont {Perdew},
  \citenamefont {Burke},\ and\ \citenamefont {Ernzerhof}}]{perdew1998perdew}%
  \BibitemOpen
  \bibfield  {author} {\bibinfo {author} {\bibfnamefont {J.~P.}\ \bibnamefont
  {Perdew}}, \bibinfo {author} {\bibfnamefont {K.}~\bibnamefont {Burke}},\ and\
  \bibinfo {author} {\bibfnamefont {M.}~\bibnamefont {Ernzerhof}},\ }\bibfield
  {title} {\bibinfo {title} {Perdew, {{Burke}}, and {{Ernzerhof Reply}}:},\
  }\href {https://doi.org/10.1103/PhysRevLett.80.891} {\bibfield  {journal}
  {\bibinfo  {journal} {Physical Review Letters}\ }\textbf {\bibinfo {volume}
  {80}},\ \bibinfo {pages} {891} (\bibinfo {year} {1998})}\BibitemShut
  {NoStop}%
\bibitem [{\citenamefont {Monkhorst}\ and\ \citenamefont
  {Pack}(1976)}]{monkhorst1976speciale}%
  \BibitemOpen
  \bibfield  {author} {\bibinfo {author} {\bibfnamefont {H.~J.}\ \bibnamefont
  {Monkhorst}}\ and\ \bibinfo {author} {\bibfnamefont {J.~D.}\ \bibnamefont
  {Pack}},\ }\bibfield  {title} {\bibinfo {title} {Special points for
  {{Brillouin-zone}} integrations},\ }\href
  {https://doi.org/10.1103/PhysRevB.13.5188} {\bibfield  {journal} {\bibinfo
  {journal} {Physical Review B}\ }\textbf {\bibinfo {volume} {13}},\ \bibinfo
  {pages} {5188} (\bibinfo {year} {1976})}\BibitemShut {NoStop}%
\bibitem [{\citenamefont {Grimme}\ \emph {et~al.}(2010)\citenamefont {Grimme},
  \citenamefont {Antony}, \citenamefont {Ehrlich},\ and\ \citenamefont
  {Krieg}}]{grimme2010consistent}%
  \BibitemOpen
  \bibfield  {author} {\bibinfo {author} {\bibfnamefont {S.}~\bibnamefont
  {Grimme}}, \bibinfo {author} {\bibfnamefont {J.}~\bibnamefont {Antony}},
  \bibinfo {author} {\bibfnamefont {S.}~\bibnamefont {Ehrlich}},\ and\ \bibinfo
  {author} {\bibfnamefont {H.}~\bibnamefont {Krieg}},\ }\bibfield  {title}
  {\bibinfo {title} {A consistent and accurate ab initio parametrization of
  density functional dispersion correction ({{DFT-D}}) for the 94 elements
  {{H-Pu}}},\ }\href {https://doi.org/10.1063/1.3382344} {\bibfield  {journal}
  {\bibinfo  {journal} {The Journal of Chemical Physics}\ }\textbf {\bibinfo
  {volume} {132}},\ \bibinfo {pages} {154104} (\bibinfo {year}
  {2010})}\BibitemShut {NoStop}%
\bibitem [{\citenamefont {Anisimov}\ \emph {et~al.}(1991)\citenamefont
  {Anisimov}, \citenamefont {Zaanen},\ and\ \citenamefont
  {Andersen}}]{anisimov1991band}%
  \BibitemOpen
  \bibfield  {author} {\bibinfo {author} {\bibfnamefont {V.~I.}\ \bibnamefont
  {Anisimov}}, \bibinfo {author} {\bibfnamefont {J.}~\bibnamefont {Zaanen}},\
  and\ \bibinfo {author} {\bibfnamefont {O.~K.}\ \bibnamefont {Andersen}},\
  }\bibfield  {title} {\bibinfo {title} {Band theory and {{Mott}} insulators:
  {{Hubbard U}} instead of {{Stoner I}}},\ }\href
  {https://doi.org/10.1103/PhysRevB.44.943} {\bibfield  {journal} {\bibinfo
  {journal} {Physical Review B}\ }\textbf {\bibinfo {volume} {44}},\ \bibinfo
  {pages} {943} (\bibinfo {year} {1991})}\BibitemShut {NoStop}%
\bibitem [{\citenamefont {Resta}(1992)}]{resta1992theory}%
  \BibitemOpen
  \bibfield  {author} {\bibinfo {author} {\bibfnamefont {R.}~\bibnamefont
  {Resta}},\ }\bibfield  {title} {\bibinfo {title} {Theory of the electric
  polarization in crystals},\ }\href@noop {} {\bibfield  {journal} {\bibinfo
  {journal} {Ferroelectrics}\ }\textbf {\bibinfo {volume} {136}},\ \bibinfo
  {pages} {51} (\bibinfo {year} {1992})}\BibitemShut {NoStop}%
\bibitem [{\citenamefont {{King-Smith}}\ and\ \citenamefont
  {Vanderbilt}(1993)}]{king-smith1993theory}%
  \BibitemOpen
  \bibfield  {author} {\bibinfo {author} {\bibfnamefont {R.~D.}\ \bibnamefont
  {{King-Smith}}}\ and\ \bibinfo {author} {\bibfnamefont {D.}~\bibnamefont
  {Vanderbilt}},\ }\bibfield  {title} {\bibinfo {title} {Theory of polarization
  of crystalline solids},\ }\href {https://doi.org/10.1103/PhysRevB.47.1651}
  {\bibfield  {journal} {\bibinfo  {journal} {Physical Review B}\ }\textbf
  {\bibinfo {volume} {47}},\ \bibinfo {pages} {1651} (\bibinfo {year}
  {1993})}\BibitemShut {NoStop}%
\bibitem [{\citenamefont {Resta}(1994)}]{resta1994macroscopic}%
  \BibitemOpen
  \bibfield  {author} {\bibinfo {author} {\bibfnamefont {R.}~\bibnamefont
  {Resta}},\ }\bibfield  {title} {\bibinfo {title} {Macroscopic polarization in
  crystalline dielectrics: The geometric phase approach},\ }\href
  {https://doi.org/10.1103/RevModPhys.66.899} {\bibfield  {journal} {\bibinfo
  {journal} {Reviews of Modern Physics}\ }\textbf {\bibinfo {volume} {66}},\
  \bibinfo {pages} {899} (\bibinfo {year} {1994})}\BibitemShut {NoStop}%
\bibitem [{\citenamefont {Feng}\ \emph {et~al.}(2024)\citenamefont {Feng},
  \citenamefont {Han}, \citenamefont {Zhang}, \citenamefont {Lin},
  \citenamefont {Gao}, \citenamefont {Yang},\ and\ \citenamefont
  {Meng}}]{feng2024van}%
  \BibitemOpen
  \bibfield  {author} {\bibinfo {author} {\bibfnamefont {Y.}~\bibnamefont
  {Feng}}, \bibinfo {author} {\bibfnamefont {J.}~\bibnamefont {Han}}, \bibinfo
  {author} {\bibfnamefont {K.}~\bibnamefont {Zhang}}, \bibinfo {author}
  {\bibfnamefont {X.}~\bibnamefont {Lin}}, \bibinfo {author} {\bibfnamefont
  {G.}~\bibnamefont {Gao}}, \bibinfo {author} {\bibfnamefont {Q.}~\bibnamefont
  {Yang}},\ and\ \bibinfo {author} {\bibfnamefont {S.}~\bibnamefont {Meng}},\
  }\bibfield  {title} {\bibinfo {title} {Van der {{Waals}} multiferroic tunnel
  junctions based on sliding multiferroic layered {{VSi$_2$N$_4$}}},\ }\href
  {https://doi.org/10.1103/PhysRevB.109.085433} {\bibfield  {journal} {\bibinfo
   {journal} {Physical Review B}\ }\textbf {\bibinfo {volume} {109}},\ \bibinfo
  {pages} {085433} (\bibinfo {year} {2024})}\BibitemShut {NoStop}%
\bibitem [{\citenamefont {Zhang}\ \emph {et~al.}(2023)\citenamefont {Zhang},
  \citenamefont {Xu}, \citenamefont {Huang}, \citenamefont {Dai}, \citenamefont
  {Kou},\ and\ \citenamefont {Ma}}]{zhang2023layerpolarized}%
  \BibitemOpen
  \bibfield  {author} {\bibinfo {author} {\bibfnamefont {T.}~\bibnamefont
  {Zhang}}, \bibinfo {author} {\bibfnamefont {X.}~\bibnamefont {Xu}}, \bibinfo
  {author} {\bibfnamefont {B.}~\bibnamefont {Huang}}, \bibinfo {author}
  {\bibfnamefont {Y.}~\bibnamefont {Dai}}, \bibinfo {author} {\bibfnamefont
  {L.}~\bibnamefont {Kou}},\ and\ \bibinfo {author} {\bibfnamefont
  {Y.}~\bibnamefont {Ma}},\ }\bibfield  {title} {\bibinfo {title}
  {Layer-polarized anomalous {{Hall}} effects in valleytronic van der {{Waals}}
  bilayers},\ }\href {https://doi.org/10.1039/D2MH00906D} {\bibfield  {journal}
  {\bibinfo  {journal} {Materials Horizons}\ }\textbf {\bibinfo {volume}
  {10}},\ \bibinfo {pages} {483} (\bibinfo {year} {2023})}\BibitemShut
  {NoStop}%
\bibitem [{\citenamefont {Li}\ \emph {et~al.}(2024)\citenamefont {Li},
  \citenamefont {Li}, \citenamefont {Lin}, \citenamefont {Zhang}, \citenamefont
  {Chen}, \citenamefont {Wu},\ and\ \citenamefont {Yang}}]{li2024sliding}%
  \BibitemOpen
  \bibfield  {author} {\bibinfo {author} {\bibfnamefont {L.}~\bibnamefont
  {Li}}, \bibinfo {author} {\bibfnamefont {X.}~\bibnamefont {Li}}, \bibinfo
  {author} {\bibfnamefont {L.}~\bibnamefont {Lin}}, \bibinfo {author}
  {\bibfnamefont {D.}~\bibnamefont {Zhang}}, \bibinfo {author} {\bibfnamefont
  {M.}~\bibnamefont {Chen}}, \bibinfo {author} {\bibfnamefont {D.}~\bibnamefont
  {Wu}},\ and\ \bibinfo {author} {\bibfnamefont {Y.}~\bibnamefont {Yang}},\
  }\bibfield  {title} {\bibinfo {title} {Sliding- and twist-tunable valley
  polarization in bilayer {{NiI}} 2},\ }\href
  {https://doi.org/10.1103/PhysRevB.110.205119} {\bibfield  {journal} {\bibinfo
   {journal} {Physical Review B}\ }\textbf {\bibinfo {volume} {110}},\ \bibinfo
  {pages} {205119} (\bibinfo {year} {2024})}\BibitemShut {NoStop}%
\bibitem [{\citenamefont {Zhong}\ \emph {et~al.}(2023)\citenamefont {Zhong},
  \citenamefont {Cheng}, \citenamefont {Ren},\ and\ \citenamefont
  {Wu}}]{zhong2023theoreticala}%
  \BibitemOpen
  \bibfield  {author} {\bibinfo {author} {\bibfnamefont {T.}~\bibnamefont
  {Zhong}}, \bibinfo {author} {\bibfnamefont {L.}~\bibnamefont {Cheng}},
  \bibinfo {author} {\bibfnamefont {Y.}~\bibnamefont {Ren}},\ and\ \bibinfo
  {author} {\bibfnamefont {M.}~\bibnamefont {Wu}},\ }\bibfield  {title}
  {\bibinfo {title} {Theoretical studies of sliding ferroelectricity,
  magnetoelectric couplings, and piezo-multiferroicity in two-dimensional
  magnetic materials},\ }\href {https://doi.org/10.1016/j.cplett.2023.140430}
  {\bibfield  {journal} {\bibinfo  {journal} {Chemical Physics Letters}\
  }\textbf {\bibinfo {volume} {818}},\ \bibinfo {pages} {140430} (\bibinfo
  {year} {2023})}\BibitemShut {NoStop}%
\bibitem [{\citenamefont {Pan}\ \emph {et~al.}(2025)\citenamefont {Pan},
  \citenamefont {Wang}, \citenamefont {Liu}, \citenamefont {Ren}, \citenamefont
  {Liu}, \citenamefont {Wang},\ and\ \citenamefont
  {Cho}}]{pan2025stackingtunable}%
  \BibitemOpen
  \bibfield  {author} {\bibinfo {author} {\bibfnamefont {Y.}~\bibnamefont
  {Pan}}, \bibinfo {author} {\bibfnamefont {C.}~\bibnamefont {Wang}}, \bibinfo
  {author} {\bibfnamefont {S.}~\bibnamefont {Liu}}, \bibinfo {author}
  {\bibfnamefont {F.}~\bibnamefont {Ren}}, \bibinfo {author} {\bibfnamefont
  {C.}~\bibnamefont {Liu}}, \bibinfo {author} {\bibfnamefont {B.}~\bibnamefont
  {Wang}},\ and\ \bibinfo {author} {\bibfnamefont {J.-H.}\ \bibnamefont
  {Cho}},\ }\bibfield  {title} {\bibinfo {title} {Stacking-tunable multiferroic
  states in bilayer {{ScI2}}},\ }\href {https://doi.org/10.1063/5.0298703}
  {\bibfield  {journal} {\bibinfo  {journal} {Applied Physics Letters}\
  }\textbf {\bibinfo {volume} {127}},\ \bibinfo {pages} {221601} (\bibinfo
  {year} {2025})},\ \bibinfo {note} {comment: 7 figures},\ \Eprint
  {https://arxiv.org/abs/2510.16379} {arXiv:2510.16379 [cond-mat]} \BibitemShut
  {NoStop}%
\end{thebibliography}%
\end{document}


\title{Supplemental Material for `` Fractional Spin Ferroelectric and Sliding Spin Current in  Magnetic Sliding Ferroelectrics "}

\author{Yilin Han}
\affiliation{Key Lab of advanced optoelectronic quantum architecture and measurement (MOE), Beijing Key Laboratory of Quantum Matter State Control and Ultra-Precision Measurement Technology, and School of Physics, Beijing Institute of Technology, Beijing 100081, China}

\author{Lei Li}
\affiliation{Research Center for Quantum Physics and Technologies, Inner Mongolia University, Inner Mongolia Key Laboratory of Microscale Physics and Atomic Manufacturing, Inner Mongolia University,\\ and School of Physical Science and Technology, Inner Mongolia University, Hohhot 010021, China}

\author{Chaoxi Cui}
\affiliation{Key Lab of advanced optoelectronic quantum architecture and measurement (MOE), Beijing Key Laboratory of Quantum Matter State Control and Ultra-Precision Measurement Technology, and School of Physics, Beijing Institute of Technology, Beijing 100081, China}

\author{Run-Wu Zhang}
\affiliation{Key Lab of advanced optoelectronic quantum architecture and measurement (MOE), Beijing Key Laboratory of Quantum Matter State Control and Ultra-Precision Measurement Technology, and School of Physics, Beijing Institute of Technology, Beijing 100081, China}

\author{Zhi-Ming Yu}
\email{zhiming\_yu@bit.edu.cn}
\affiliation{Key Lab of advanced optoelectronic quantum architecture and measurement (MOE), Beijing Key Laboratory of Quantum Matter State Control and Ultra-Precision Measurement Technology, and School of Physics, Beijing Institute of Technology, Beijing 100081, China}

\author{Yugui Yao}
\affiliation{Key Lab of advanced optoelectronic quantum architecture and measurement (MOE), Beijing Key Laboratory of Quantum Matter State Control and Ultra-Precision Measurement Technology, and School of Physics, Beijing Institute of Technology, Beijing 100081, China}

\maketitle
\section{Calculation Methods}
The first-principles calculations of these magnetic sliding ferroelectrics (SFEs) were performed using the Vienna ab initio simulation package (VASP) \cite{kresse1996efficient,kresse1996efficiency}. The projector augmented wave (PAW) pseudopotentials \cite{blochl1994projector,kresse1999ultrasoft} were adopted in the calculation and generalized gradient approximation (GGA) \cite{perdew1996generalized} of the Perdew-Burke-Ernzerhof (PBE) \cite{perdew1998perdew} functional was selected as the exchange-correlation potential. The 2D Brillouin zone (BZ) was sampled using Monkhorst-Pack k-mesh \cite{monkhorst1976speciale} with a size of $21\times21\times1$. The energy cutoff was set to 600 eV and the energy convergence criteria was $10^{-6}$ eV. DFT-D3 method \cite{grimme2010consistent} was used to consider the van der Waals correction in all of our bilayers. Since all bilayers considered in this work contain transition-metal elements, the on-site Hubbard correction was included via the PBE+$U$ method \cite{anisimov1991band}. The effective Hubbard $U$ values applied to Cr atom in CrI$_3$ and V atom in V$X_2$ ($X=$ S, Se, Te) were both set to 3 eV.
The optimized in-plane lattice constants of bilayer CrI$_3$, VS$_2$, VSe$_2$, and VTe$_2$ are approximately 6.87 \AA, 3.22 \AA, 3.32 \AA, and 3.53 \AA, respectively.

The electronic polarization and spin electronic polarization were obtained by post-processing the output of VASP calculations, where $e$ was taken as the electronic charge, namely $e=-1.6\times10^{-19}$ C. Owing to the two-dimensional nature of these bilayers, the in-plane polarization was evaluated using the modern theory of polarization \cite{resta1992theory,king-smith1993theory,resta1994macroscopic,vanderbilt2018berry}, whereas the out-of-plane polarization was calculated from the dipole moment.

\section{Intermediate state of bilayer $2H$-V$X_2$}
\begin{figure}[!htbp]
\centering
\includegraphics[width=0.8\textwidth]{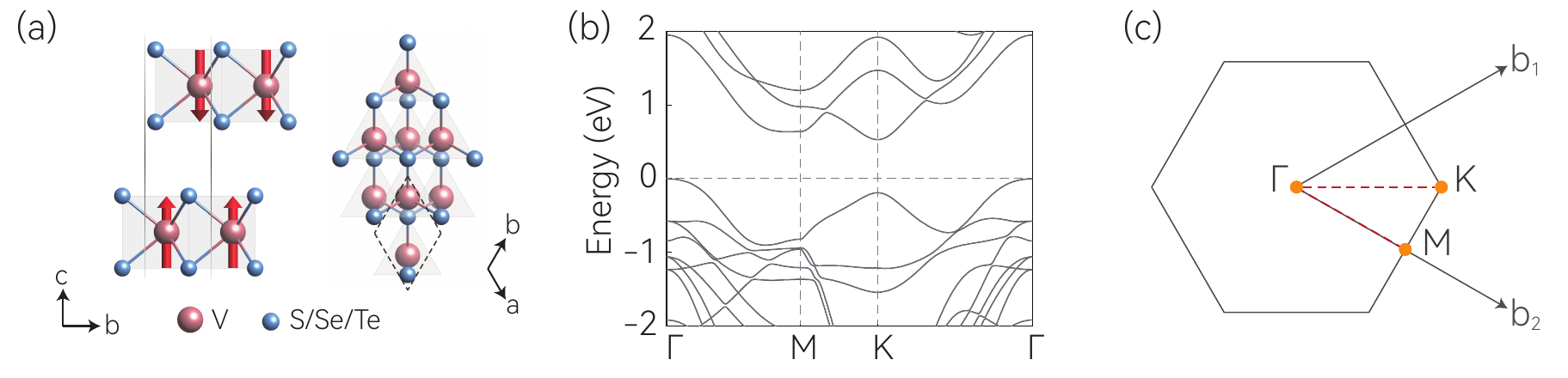}
\caption{\label{fig:IS} (a) Side and top views of bilayer $2H$-V$X_2$ in intermediate state (IS). (b) Band structures of bilayer VS$_2$ in IS stacking, which is a type-IV magnet with spin-degenerate bands. (c) The two-dimensional BZ of bilayer V$X_2$.}
\end{figure}

The atomic and magnetic structures of the intermediate-state (IS) stacking of $R$-stacked bilayer $2H$-V$X_2$ are shown in Fig.~\ref{fig:IS}(a).
In this configuration, the V atoms with opposite magnetic moments are related by $\{{\cal M}_z{\cal T}|\frac{1}{2}\frac{1}{2}0\}$ symmetry, indicating the IS stacking a type-IV magnetic state.
Accordingly, the calculated electronic bands are spin degenerate, as shown in Fig.~\ref{fig:IS}(b).
The corresponding two-dimensional Brillouin zone (BZ) and high-symmetry path are illustrated in Fig.~\ref{fig:IS}(c).

\section{Other magnetic SFEs for sliding spin current}
\begin{figure}[!htbp]
\centering
\includegraphics[width=0.8\textwidth]{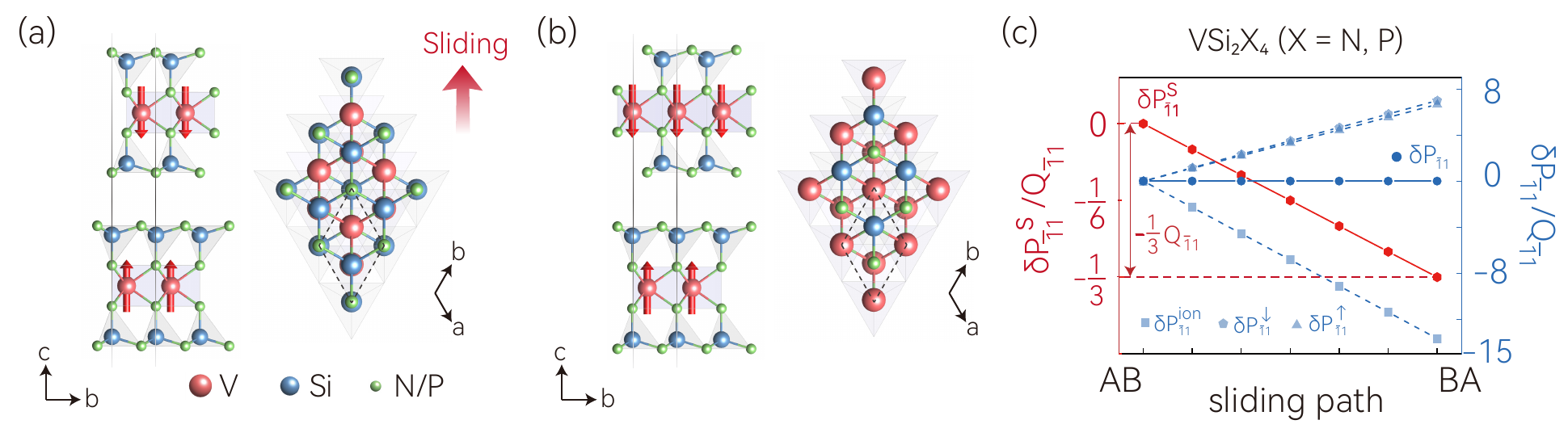}
\caption{\label{fig:VSi2N4} (a-b) Side and top views of bilayer VSi$_2X_4$ ($X=$ N, P) in (a) AB stacking, and (b) BA stacking. (c) Variations of electronic polarization and spin electronic polarization along the ($\bar 1$1) direction.  $Q_a=1110.3$ pC/m ($926.1$ pC/m) denotes the polarization quantum along ($\bar{1}1$) direction for VSi$_2$N$_4$ (VSi$_2$P$_4$).}
\end{figure}

\begin{figure}[!htbp]
\centering
\includegraphics[width=0.75\textwidth]{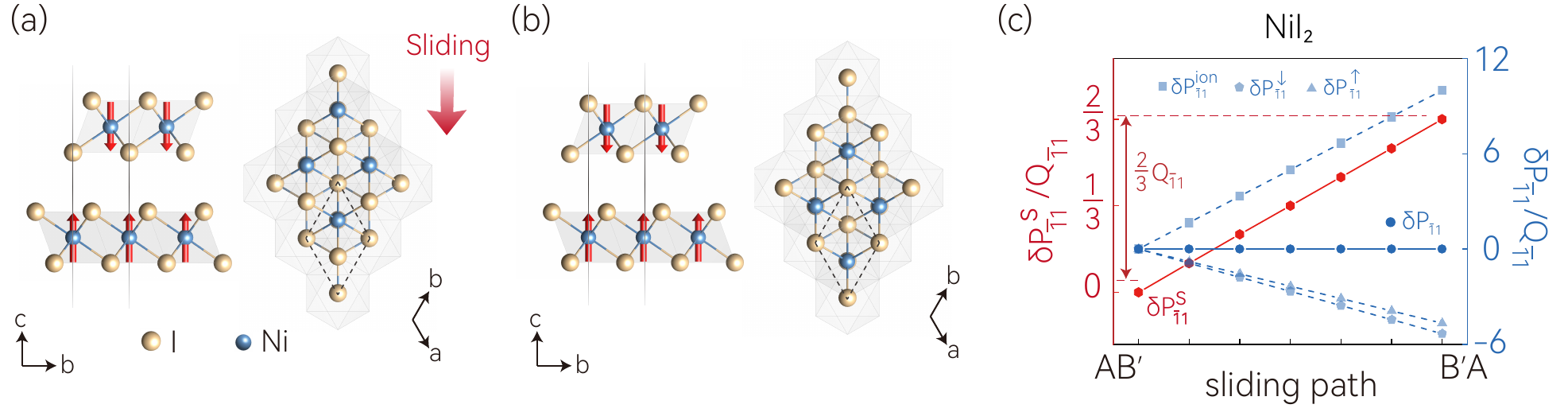}
\caption{\label{fig:NiI2} (a-b) Side and top views of bilayer NiI$_2$ in (a) AB' stacking, and (b) B'A stacking. (c) Variations of electronic polarization and spin electronic polarization along the ($\bar 1$1) direction.  $ Q_{\bar{1}1}=817.4$ pC/m denotes the polarization quantum along ($\bar{1}1$) direction.}
\end{figure}

\begin{figure}[H]
\centering
\includegraphics[width=0.9\textwidth]{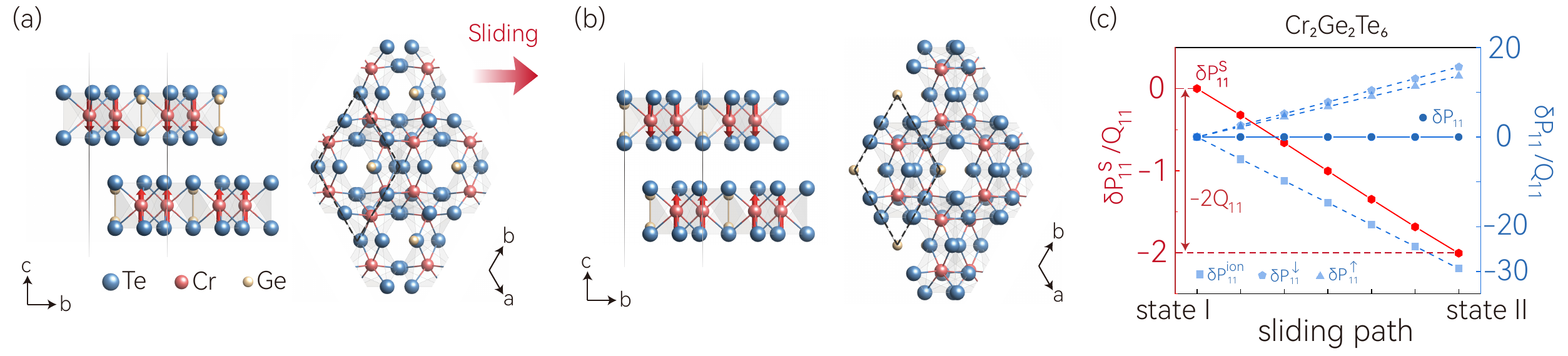}
\caption{\label{fig:CrGeTe} (a-b) Side and top views of bilayer Cr$_2$Ge$_2$Te$_6$ in ferroelectric (a) state I and (b) state II. (c) Variations of electronic polarization and spin electronic polarization along the (11) direction.  $ Q_{11}=270.5$ pC/m denotes the polarization quantum along (11) direction.}
\end{figure}

\begin{figure}[H]
\centering
\includegraphics[width=0.8\textwidth]{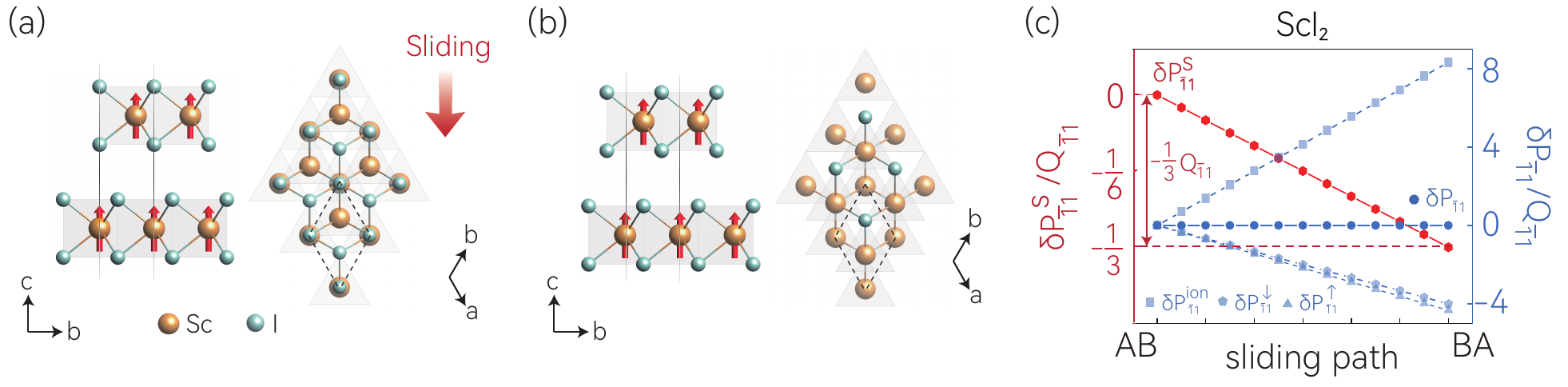}
\caption{\label{fig:ScI2} (a-b) Side and top views of bilayer ScI$_2$ in (a) AB stacking, and (b) BA stacking. (c) Variations of electronic polarization and spin electronic polarization along the ($\bar 1$1) direction.  $ Q_{\bar{1}1}=783.3$ pC/m denotes the polarization quantum along ($\bar{1}1$) direction.}
\end{figure}

To further demonstrate the generality of the sliding spin current proposed here, we performed calculations for several additional magnetic bilayers, including the compensated ferrimagnetic (cFiM) bilayers V$X_2$ ($X=$ Se, Te), VSi$_2X_4$ ($X=$ N, P), NiI$_2$, Cr$_2$Ge$_2$Te$_6$, as well as the ferromagnetic (FM) bilayer ScI$_2$. All these systems have been previously identified as SFEs with low sliding barriers, indicating that ferroelectric switching is energetically feasible. \cite{ma2024strain,sheng2025ubiquitous,feng2024van,zhang2023layerpolarized,li2024sliding,zhong2023theoreticala,pan2025stackingtunable}. For VSe$_2$ and VTe$_2$, the atomic structures and polarization evolution along the sliding path have been discussed in the main text, where the variation of polarization are consistent with those of VS$_2$ after accounting for the polarization quantum $Q_{\bar11}$ [see Figs.~4(a-d)]. For other magnetic SFEs, the atomic structures of the two endpoint states, together with the associated variations of ${\bm P}^{\rm ion}$, ${\bm P}^{\uparrow}$, ${\bm P}^{\downarrow}$, ${\bm P}$ and ${\bm P}^{\rm s}$  along the sliding paths, are shown in Figs.~\ref{fig:VSi2N4}--\ref{fig:ScI2}. In all these cases, $P_{\parallel}$ remains nearly unchanged along the sliding pathway, whereas the spin electronic polarization $P^s_{\parallel}$ exhibits a finite change, demonstrating the generation of a pure spin current during ferroelectric switching.

\FloatBarrier
\bibliography{mybib-SM}